# Enhancing Interaction with Augmented Reality through Mid-Air Haptic Feedback: Architecture Design and User Feedback

**Diego Vaquero-Melchor *,† and Ana M. Bernardos †**

Information Processing and Telecommunications Center, Universidad Politécnica de Madrid, 28040 Madrid, Spain; abernardos@grpss.ssr.upm.es

* Correspondence: diego.vaquero@grpss.ssr.upm.es

† Current address: ETSIT, Av. Complutense 30, 28040 Madrid, Spain.

**Abstract:** Nowadays, Augmented-Reality (AR) head-mounted displays (HMD) deliver a more immersive visualization of virtual contents, but the available means of interaction, mainly based on gesture and/or voice, are yet limited and obviously lack realism and expressivity when compared to traditional physical means. In this sense, the integration of haptics within AR may help to deliver an enriched experience, while facilitating the performance of specific actions, such as repositioning or resizing tasks, that are still dependent on the user's skills. In this direction, this paper gathers the description of a flexible architecture designed to deploy haptically enabled AR applications both for mobile and wearable visualization devices. The haptic feedback may be generated through a variety of devices (e.g., wearable, graspable, or mid-air ones), and the architecture facilitates handling the specificity of each. For this reason, within the paper, it is discussed how to generate a haptic representation of a 3D digital object depending on the application and the target device. Additionally, the paper includes an analysis of practical, relevant issues that arise when setting up a system to work with specific devices like HMD (e.g., HoloLens) and mid-air haptic devices (e.g., Ultrahaptics), such as the alignment between the real world and the virtual one. The architecture applicability is demonstrated through the implementation of two applications: (a) Form Inspector and (b) Simon Game, built for HoloLens and iOS mobile phones for visualization and for UHK for mid-air haptics delivery. These applications have been used to explore with nine users the efficiency, meaningfulness, and usefulness of mid-air haptics for form perception, object resizing, and push interaction tasks. Results show that, although mobile interaction is preferred when this option is available, haptics turn out to be more meaningful in identifying shapes when compared to what users initially expect and in contributing to the execution of resizing tasks. Moreover, this preliminary user study reveals some design issues when working with haptic AR. For example, users may be expecting a tailored interface metaphor, not necessarily inspired in natural interaction. This has been the case of our proposal of virtual pressable buttons, built mimicking real buttons by using haptics, but differently interpreted by the study participants.

## 1. Introduction

In recent years, Augmented Reality (AR) has become a popular technology due to its integration in mobile apps (e.g., the Pokemon Go game [1]) and the increasing availability of powerful devices and development frameworks that enable the deployment of specialized solutions (e.g., for maintenance, training, or surveillance).

In its most basic definition, augmented reality consists of combining digital information superimposed on a view of the real world [2]. However, this definition only emphasizes the visualization component, but sound interaction methods for object manipulation and task completion are also key for AR service delivery. In case of handheld devices, such as smartphones, the interaction is usually performed through touch control, while in Head-Mounted Display (HMD) alternatives, the interaction is in general performed through gestures, voice, or a combination of both or by using physical controllers. Actually, there is no dominant design yet for HMD interaction: although these devices enable more natural space exploration, real-world interaction metaphors cannot be directly used. Moreover, wearable AR (AR enabled through headsets/glasses as visualization tools) share the alignment challenge with mobile-enabled AR. This refers to the perception problems due to misalignment of the virtual content with respect to the real world, together with other issues that may affect the level of realism or the lack of precision for object/scene manipulation.

In this paper, we explore the integration of haptic feedback as a way to enhance perception and manipulation of AR virtual assets. The contribution of the paper is threefold. First, we analyze how a 3D virtual object can be haptically represented to be perceived in shape and volume (aspects such as weight, texture, or temperature are not considered due to technology limitations), proposing some different representation options valid for different haptic approaches: mid-air, vibratile, or touchable. Next, we propose an architecture that is designed to facilitate the development of applications combining both AR and haptics (which we refer as *harptic*), on different types of visualization and interaction devices. This architecture proposes a data model and a workflow designed to be easily extensible and adaptable to new platforms as well as a series of artifacts ready to be integrated in future deployments. Overall, the main objective is to implement a natural interaction experience both when using mobile and wearable devices. Mobile devices can be integrated in the architecture either as visualization or haptic interactive devices. Moreover, the flexibility of the architecture is that the generation of haptic feedback can be adapted to any type of device. In addition to allowing the user to move in the real world while using them, mobile devices also enable to generate, e.g., vibratile haptic feedback, which may be integrated with AR experiences. For example, vibration may be used to show contact between the AR content and the device itself (e.g., the user is "touching" the 3D model with the device). Another use could be, for example, applying different vibration patterns to communicate a certain texture to the user. To demonstrate the feasibility of this architecture, we propose its validation with the construction of two proof-of-concept applications: (i) *Shape Inspector*, an application that generates 3D haptic shapes, with and without visual support for the user to identify them, and (ii) a *Simon-like game* [3], in which a system with four virtual coloured buttons is built and the user is asked to repeat a sequence of colors by virtually pressing them. Both prototypes combine HoloLens and mid-air haptics (i.e., hand-based haptic experience without contact surface enabled in this case by ultrasounds (Ultrahaptics device, UHDK5 version). Finally we expose the results of the user tests carried out on the applications mentioned before; the trial objective, beyond the architecture feasibility demonstration, is focused on analyzing the value of mid-air haptics to improve visualization: object guessability, perception, or resizing tasks are considered. In spite of the inherent technology limitations, 3D object implementation over mid-air haptics may be in practice more expressive than initially envisioned by the users.

The rest of this paper is structured as follows. Section 2 gives an overview of the current state of the integration of augmented reality in relation with mid-air haptic devices as well as an overview of the associated problem. After that, in Section 3, we discuss how a 3D virtual object can be haptically represented. Later, in Section 4, taking into account the solutions proposed in the literature and the challenges exposed in the state-of-the-art, we present our design of a scalable, platform-independent architecture that allows an easy integration of mid-air haptic feedback in AR context. Section 5 contains a review of the main practical issues encountered during the development stage, in particular those related with visual alignment and the apparition of non-desired haptic feedback. After that, in Section 6, we propose two proofs of concept that validates the proposed architecture. Section 7 details the results of the user tests. Results show that the proposed architecture enables to deploy applications which combine different technical objectives on different devices. Besides that, results show that haptic feedback may be useful as support for working with AR content. Next, in Section 8, the experimental results are presented. Finally, Section 9 contains the main conclusions and future research lines.

## 2. State of the Art

When interacting with computers or machines, *natural interaction* usually refers to the final goal of the user experience that aims at being smooth, easy, and evident, i.e., natural to the user. However, natural interaction is a subjective concept [4] that can be applied to a wide range of fields. This is mainly due to different users having different preferences, expectations, and background. Several studies have been conducted regarding the interaction with virtual content without hands, that is, without direct contact from the user. Some approaches [5,6] have made use of users' gaze to show or hide information when using AR displays. It has also been studied whether the use of head gazing or eye pointing [7] is more effective when interacting. Results indicate that users tend to prefer the head approach because of its better stability. Given that this type of interaction is quite widespread, some frameworks such as EyeMRTK [8] have been appearing to facilitate developer tasks. The survey by Mohamed Khamis et al. [9] collects the history and current situation of gaze-enabled interaction in the context of mobile handheld devices. Finally, a survey about interaction methods for head-mounted displays that emphasizes touch and touchless-based interaction can be found in Reference [10]. Regarding direct manipulation, users may apply actions over elements like "touching" or "moving". In the end, the better the interaction adapts to what the user expects, the more natural it will feel. Hence, different systems and devices may offer different natural ways to interact with them. In wearable AR context, the use of hand gestures has been widely explored [11,12]. Thanks to the apparition of more powerful devices, nowadays, it is possible not only to detect gestures but also to perform image analysis techniques in real time and to track hand movements (e.g., Kinect and Leap Motion [13,14]). The preference of users for certain gestures has already been explored in previous works [15]. Nevertheless, the interaction with the virtual content is not natural yet. This is mainly due to one key component that has not been systematically integrated into this interaction: tactile feedback. The concept of haptic feedback promises to tear down the boundaries between the real, physical world and the virtual one. The combination of hand gestures and haptic feedback is positioned as a middle ground between precision and comfort, and its advantages have already been explored in the literature [16].

On its most basics, haptic feedback consists of presenting synthetic tactile stimuli to the users. Since this type of feedback can be generated on demand, it can be used to enhance the perception of the AR content. Whereas before the user could only see a digital object, now they can touch it or perceive it. There are many different types of haptic devices available today. Two main groups can be distinguished depending on whether the feedback is based on the perception of forces (*kinesthetic*) or produced in the skin of the user (*cutaneous*, tactile if the perception is over the fingers).

The first group is formed by kinesthetic devices. This kind of device tends to present low portability because they are grounded. Some examples of these devices are the one used in the PHANToM project [17]—which consists of a grounded pen attached to a device that generates haptic feedback—or the one proposed in Reference [18]—a grounded, graspable haptic device that kept track of the user's hand and generated the feedback, e.g., to simulate contact with digital elements.

The second group is composed of cutaneous devices, and it includes devices such as wristbands [19–21], haptic rings [22,23], and gloves [24–26] or mid-air tools. More in detail, cutaneous devices can be divided into the following groups:

- *Basic cutaneous devices*. This kind of device is one of the most popular, since they are simple and provide great results. The vibration of the device is perceived by the user through cutaneous perception that the user is accustomed to, such as when they receive an alert on the smartphone. However, this kind of device presents some limitations; for example, users may find it difficult to understand complicated vibration patterns. Some examples of this kind of device are vibration wristbands [19] (discontinued from 2018) [27]. Also, there are possibilities like fingertip devices [28] or smartphone vibration itself.
- *Active surfaces*. This devices uses physical components to represent the feedback, for example, by using vertical bars. This kind of device is good for rendering surfaces, having nice resolution and accuracy. However, they have limited functionality due to the actuators required to perform the feedback. The biggest limitation of this kind of device is the lack of portability, since they are usually required to be installed in a certain manner. Some examples of this kind of device are Relief project [29] and Materiable [30].
- *Mid-air* devices provide haptic feedback without requiring the user to wear any device (in contrast with devices such as gloves, bracelets, or rings). The main difference with the previous group is that, in the previous one, the user came into physical contact with the haptic device, while in this group, that interaction occurs in the air (hence, the name). There are different kinds of mid-air haptic devices, but they can be divided into three groups according to the technology they use: air vortices [31], air-jets [32], and ultrasounds [33,34].

Ultrasound-based devices use ultrasonic speakers to focus sound waves at a point in the space above them. At the point where the waves converge, a pressure point is generated. That pressure point is strong enough to be noticed in the user's hand and, thus, to induce a tactile sensation. By changing the frequency of the ultrasounds, they are also able to control the strength of the sensation. A representative of this type of device is Ultrahaptics [33,35]. Some works have already explored the idea of using those haptic points to render 3D objects [36]. In this case, to render means to dispose of the haptics points and to let the user perceive them with her hands.

Some authors justify the use of technology based on air-vortex-ring generators. Shtarbanov and Bove [37] presented an approach to augment 2D and 3D displays with haptic capabilities. Their reasons for the justification were the limitations that ultrasound presented for their system: weak perceived sensation, limited range, and volume of interaction limited by the area of the array.

The use of haptics to interact with digital assets has been in literature for HMD [38], physical traditional screens [39], and custom hardware solutions [40,41]. In comparison, the idea of combining AR and MR with mid-air haptic feedback has not been so thoroughly explored. Moreover, literature related with this topic tends to be focused on the application stage. Some works are only focused on the haptic side of the equation, such as the work of Leithinger [42]. In his work, he replaced visual aids with haptic feedback using Ultrahaptics UHDK5 device. These sensations that surround the hand like a haptic aura guided the user towards a physical control element with the objective to provide information about its functionality and type of control. Briefly, they made use of haptic feedback to provide information about a physical real-world element. In the same line of not incorporating AR information, Goncu and Kroll [43] presented the GraVVITAS system, a multi-modal system for presenting 2D graphics. The system they exposed had the limitation of not being able to display 3D graphics, and to solve this problem, they extended the system to integrate the Ultrahaptics UHEV1 device. In that system, a 3D model was generated using an external tool. In the next step, that 3D model was divided into slices according to its volume. The result of that slicing process was then haptically presented to the user by selecting the slice associated to the user's palm position. From this work, two main limitations can be obtained. First, their system had the limitation of requiring a specific program for the generation of the 3D model, OpenSCAD. Second, their system was limited to iOS and MacOS devices, without offering possibilities for the integration of more devices. Other authors have created interaction metaphors to generate tools to manipulate digital elements using mid-air devices [44]. In Geerts's work, he [45] goes one step further and removes the physical component of the interaction. Instead of having a real button, the element is presented as a haptic, augmented-reality button to the user. His work explores the idea about how to design visual and haptic interaction patterns. This work demonstrates that a physical component can be simulated through haptic interaction and that users are really able to interact with them. Dzidek et al. [46] combined AR with mid-air haptics and conducted an evaluation with users. Their focus is on how gestures and actions are performed over the haptic device. Results show that the improved graphic components with haptic rendering happened to be more memorable. For the authors, this suggests that active exploration of haptic and AR content may be more convenient rather than in a purely receptive manner. This same train of thought has been followed in our investigation. Going forward with what will be indicated later, one of the things we seek is to know how the user explores AR models enhanced with haptic. Frutos-Pascual et al.[47] conducted an experiment to analyze the influence of mid-air haptic feedback as supplementary feedback in Virtual Reality (VR) context. They found that visual feedback was preferred among the users as supplementary cue but that, actually, haptics did increase the accuracy when dealing with small targets. These results agree with those obtained through our validation with users. In one of the tasks (resizing an AR model with haptic clues), although the users had visual feedback, the haptic one has proved more useful in reducing errors.

Mid-air haptic technologies can also be combined with gesture recognition techniques to generate attractive and powerful results. Kervegant et al. [48] briefly explored the idea of incorporating mid-air haptic feedback to the AR experience using HoloLens and Ultrahaptics. Their work consisted of displaying a virtual sphere over an Ultrahaptics device with which the user can interact. That interaction was limited to exploring the collisions between the user's hand and the sphere. As a summary, we can draw attention over the following challenges. First, a standard way to generalize the development of applications that combine mid-air haptic interaction and AR has not been proposed. So far, each system creates an ad hoc solution. This is in part due to each AR and haptic device making use of its own development platform, which makes it difficult to establish a common framework. Second, at this moment, the generation of haptic feedback associated with AR content and its full potential has not been deeply explored.

Based on these challenges, we propose an architecture that facilitates the generation and deployment of AR-haptic systems. First, it has been designed taking into account the hardware limitations of current haptic devices and that, in the future, new devices with greater potential may appear. Therefore, the architecture will be able to integrate those in a seamless way. Second, the architecture itself propitiates a workflow that encompasses the process of creating applications that combine AR and haptics. Moreover, as a result of implementations developed on the architecture, we have created a series of artifacts that can be easily integrated into future developments or adapted to new platforms. Finally, the data model that the architecture manages can be extended to more types of haptic devices, without this entailing many problems.

In this paper, we will take as an example haptic mid-air devices, which among other features do not require the user to wear anything and can be arranged on any surface. Later, we will introduce the use of the Ultrahaptics device as a representative of this group, which will be the one used to implement the prototypes that validate the architecture. In the next section, first, the concept of haptic representation of a 3D object using mid-air haptic devices and its associated problems are presented. From there, our architecture for AR-haptic (*HARptic*) interaction is exposed in detail.

## 3. Haptic Representation of an Object's Shape and Size by Using Haptic Devices

A first issue to face when defining the integration of AR and haptics is how to make a sound and expressive haptic representation of a digital element or object. Ignoring the limits imposed by technology, ideally the haptic representation of that element should coincide with the "physical" characteristics of the object (i.e., size, volume, textures, weight, etc.). This kind of representation is specially adapted to mid-air devices and gloves, that is, those devices in which the user receives the feedback in their hand. For example, if a user is haptically inspecting a 3D model of a soda bottle, they should be able to perceive with their hands the physical features of the real bottle from which the 3D model was extracted. In reality, this interpretation presents several limitations. The most obvious issue is related with the accuracy and capabilities of the haptic device itself.

In general mid-air devices deliver haptic feedback as points located over the own device. This happens in both ultrasonic (e.g., Ultrahaptics) and air-vortex-ring based devices. Hence, that point is the basic representation unit. Due to the technology in use to provide the feedback, points may not be perceived by the user as a sharp point. Instead, it may perceived more as an air field centered on that point with a diffuse radius. Figure 1 illustrates this issue. In the left side of the figure appears a "sharp" point with well-defined limits. In the right side, a situation similar to the one perceived with the Ultrahaptics is represented [33].

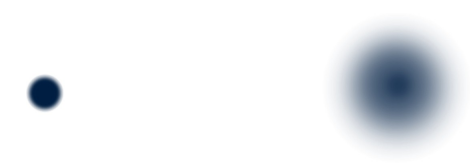

**Figure 1.** Diffuse point problem in mid-air haptic systems.

The objective is to combine several of these *fuzzy* haptic points to generate the haptic representation of an object. We will assume that any digital object may be represented as a 3D model, i.e., as a representation by means of vertices and faces. Thus, an ideal figure such as a sphere (which by definition has no vertices) becomes a representation based on adjacent faces composing that sphere. Depending on the haptic experience to be provided due to the service goal, it is feasible to generate different representations of the same object. We classify the representations as follows:

- High-fidelity representation: This type of representation seeks to give haptic feedback that resembles as close as possible the physical characteristics of the object to be represented. More specifically, the characteristics that we contemplate in this paper are volume, size, and shape, leaving aside more advanced ones like weight and textures due to technology limitations. Hence, the haptic device will need to reproduce the 3D model characteristics, mainly the vertices and the faces. This type of representation is useful in those situations in which it is required for the user to explore a model (e.g., to perceive its dimensions, to resize it, etc.). To build high-fidelity representations, it is feasible to use different strategies:
  - *Vertex-based* represents only the vertices that shape the model as scattered points.
  - *Edge-based* represents only the edges between the devices of the model. These edges are defined by the faces. By definition, vertices are also represented, since they are present in the limit of the edges.
  - *Volume-based*: In this case, both vertices and edges are represented as are the surface of the faces as well as the interior of the figure.
- Single-feature-based representation: In this case, we seek to approximate a 3D object to one or several haptic points, leaving aside fidelity. This is especially appropriate for those 3D models that are small enough and in which details would go unnoticed when using the high-fidelity representation. Instead of representing all the features that form the 3D model, it may be translated to a single point, calculated, for example, using the centroid of all its vertices. Figure 2 illustrates the issue in which the larger the 3D object is, the worse the adaptation will be using this type of representation.

To illustrate these two possibilities in practice, in Section 6, two prototypes that make use of high-fidelity and single-feature-based representation are presented. The first prototype is the *Haptic Inspector*, which uses the *high-fidelity representation* in its *edge-based* form, since we want to allow the user to explore the details of a 3D model as accurately as possible (within the limitations of the technology). However, in the *Simon* [3] prototype, we are interested in haptically representing where the colored buttons of the game are and in allowing the user to interact with them. The details of its shape are not so important, so they stay in the background. Hence, we use the *feature-based representation* for this implementation. This whole way of understanding haptic representation is also applicable to devices with vertical physical feedback such as the Materiable system [30]—a two-dimensional array of physical bars; hence, it is not limited to mid-air haptics.

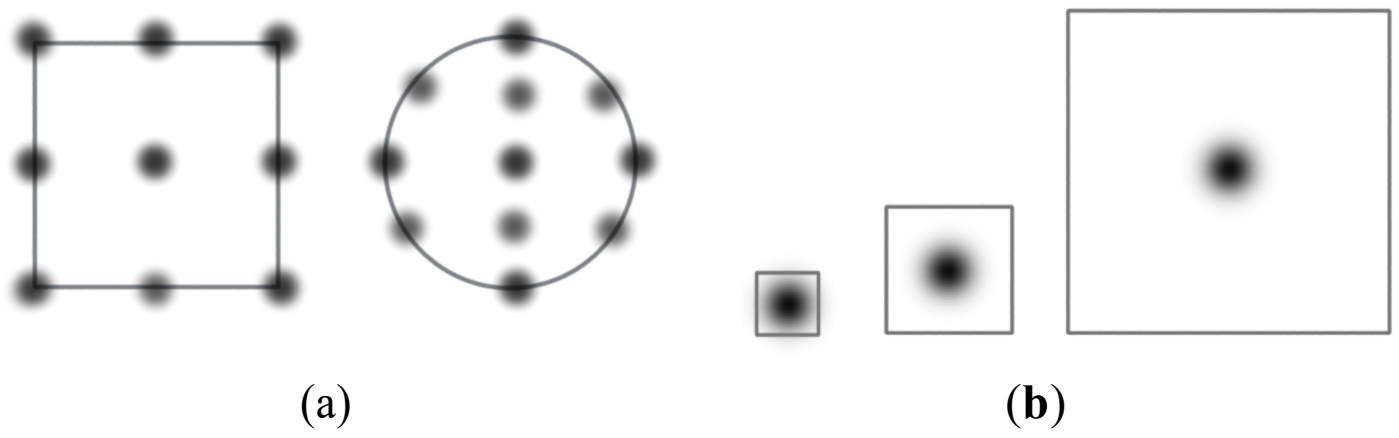

**Figure 2.** High-fidelity and feature-based haptic representations of a model: (**a**) High-fidelity representation; (b) feature-based representation.

## 4. HARP: An Architecture for Haptically Enhanced AR Experiences

Up to now, building AR applications enriched with mid-air feedback and interactivity requires a tailored development, which is constrained by the device to be used, i.e., its main features and capabilities. Additionally, in most cases, the AR experience to deliver implies coordinating the haptic representation with certain visual information and interactive capabilities. The goal of our proposal is to enhance the perception of AR 3D content, while facilitating the generation of applications and allowing various types of haptic devices to be integrated. Figure 3 shows in a nutshell a scheme of the objective pursued by the architecture. The system built over the architecture will need to support the following high-level requirements:

- The solution should be able to generate a haptic representation for **any 3D model** (taking into account the limitations of the technology in use), and it should be able to work with both haptic representation possibilities, *high-fidelity* and *feature-based*. Other types of representations should be easily incorporated into the system.
- The haptic representation of a given virtual object must be **aligned with the visual representation** of the digital element. In other words, given a 3D model with a certain appearance located in a certain position with respect to the real world, its haptic representation must occupy the same space as it while adapting to its form.
- The solution must be **multi-platform**. Since each AR device works with its own platform and development frameworks, it is necessary to define a workflow to support all of them and to speed up the development process.
- The objective is to allow the integration of **any type of haptic device**, be it mid-air, a glove, or a bracelet. As a starting point and to show the viability of the solution, it should support the Ultrahaptics device, which we use as a representative of mid-air haptic devices. However, it should be extensible to other kind of haptic devices. The main reason for focusing on mid-air technology is that it does not require the user to wear any device. Additionally, the availability of some recent devices and integrable technology that might be available as part of commercial equipment makes it interesting to explore both the performance and acceptance of mid-air devices.
- To maximize compatibility with other platforms, the architecture should work with structures similar to those of the **main development frameworks**.

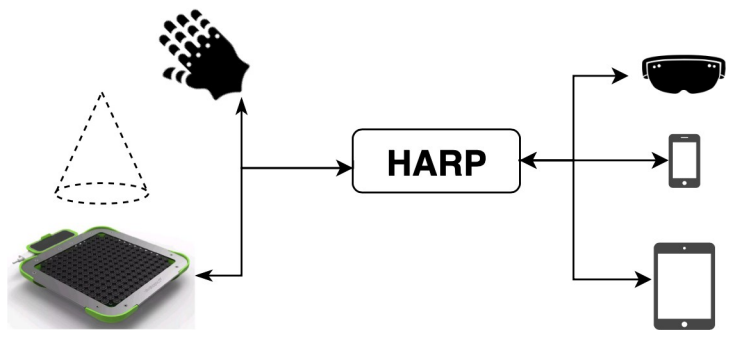

**Figure 3.** Approach to the HARP solution: haptic experience aligned with augmented reality object models.

Next, in Section 4.1, we first present the data model of the architecture, and then, in Section 4.2 the architecture itself is presented.

### 4.1. Data Model

A pillar of the proposed architecture is its data model. It is built on an adaptation of the one used by most 3D development tools, so it allows a great flexibility. When integrating the HARP architecture in different platforms, the translation from their inner representation to our proposal will be practically direct. The key concepts on which the data model is built are *sessions* and *nodes*. Figure 4 shows a class diagram with the main entities that the system manages. A *session* has users and devices connected to it (Figure 4). A device will be each client that participates in the process of augmenting AR through haptics. For example, one client may represent the application that acts as a bridge with the haptic device and another client may be the one with AR capabilities. Later on, when the components of the architecture are presented, this concept will berevisited.

A *session* has a *session status* associated which just has a reference to a root *node*, to which the remaining ones will be attached. Each *node* will have a *mesh* and a *transform* as well as a reference to the parent node and to its children. The *transform* will store the position, rotation, and scale of the node, and it is an abstraction of the same concept that the main tools and Software Development Kits (SDKs) use. For its part, we have decided to summarize the *mesh* in the minimum possible components: the vertices that form the volume, the normals for each vertex, and the triangles between them. It would be possible to add more information here, but we have found that, with these few concepts, the architecture is viable. Facing other types of haptic devices, for example, an armband or graspable ones (e.g., smartphones), the data model can be applied with some considerations. This kind of device does not display spatial haptic information, but they present a pulse of haptic feedback for a certain period of time with a certain intensity. How could a haptic figure be represented on these devices? A possibility is to approximate the entire 3D model to a single point (feature-based representation) and, under certain conditions (e.g., the user's hand or the device occupies the same position as the element), the haptic feedback is activated. As it can be seen, the data model itself is decoupled from the type of haptic device in which the representation will be performed.

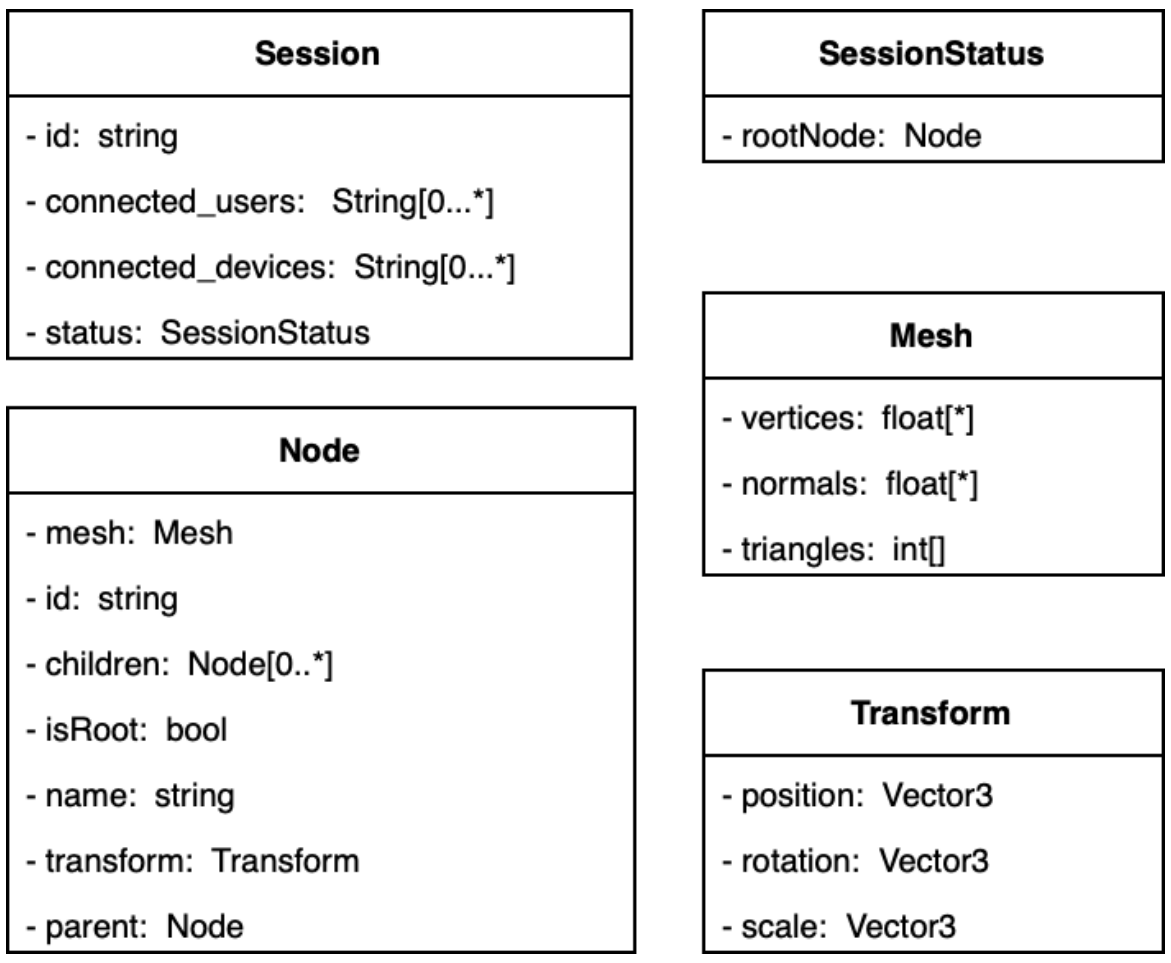

**Figure 4.** Basic data model handled by the system.

### 4.2. System Architecture

Once the data model has been presented, it is time to present the architecture itself. The diagram in Figure 5 shows the main components of the architecture, which can be divided in two main blocks: (a) the *HARP client* and (b) the *communication service*.

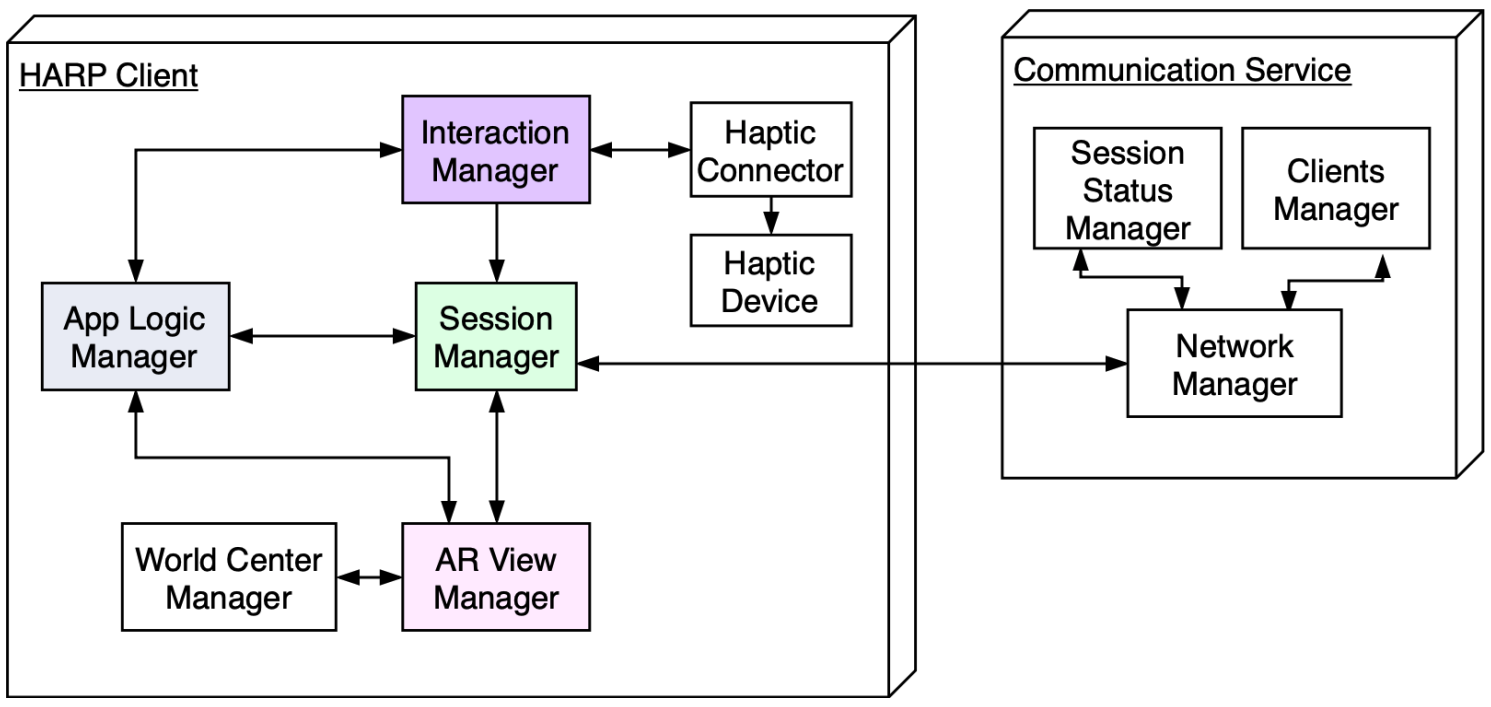

**Figure 5.** Blocks that make up the architecture with minimum possible deployment.

- HARP client: This block brings together the minimum functionalities that a client must accomplish to satisfy the requirements mentioned in Section 3:
  - The *Session Manager* component acts as an entry point into the system. As its name indicates, it is in charge of managing the *session* information, which, briefly recalling, includes information of the 3D objects to be represented by combining AR and haptics. It is capable of receiving the status of the session from the *Communication Service* and of generating and updating it.
  - The *AR View Manager* has a very specific purpose. It is in charge of receiving the *session status* from the *Session Manager* and representing it using the AR framework of the device on which it is deployed. For example, in the case of being deployed to an iOS device, after receiving the *session status*, it has to generate the scene and to create a node (SCNNode in Scenekit) on that scene for each element on the *session*. As updates from the state arrive, it must adapt the AR view to them. At this point, an issue arises: to which point in the physical space should that AR content be attached?
  - The *World Center Manager* component emerges to address that issue. It must define the required functionality to set the origin of coordinates for the digital content and its relation to the real world. The final implementation is open to all kind of possibilities. In Section 5.1, we give as an example the possibility of establishing the alignment manually or semiautomatically.
  - The *App Logic manager* groups all the logic on which the application depends. For example, in the Simon example in Section 6.2, this component was integrated on the HoloLens client. Its main occupation was to keep track of the buttons that the user presses, the generation of new color sequences, and other related functionality.
  - The *Interaction Manager* component abstracts the interaction capabilities of the system. For example, the Simon prototype was deployed in two different devices: an iOS one and an HoloLens one. On the one hand, the *Interaction Manager* from the iOS client is in charge of detecting touches on the screen, translating them into touches on digital objects and generating the corresponding events. On the other hand, the implementation on the HoloLens client receives information of hand gestures and reacts to it. In any case, the resulting interaction should be able to trigger certain logics of the application and to cooperate with the *Session Manager*. An example of this process in the Simon system happens when the user pushes the haptic button with their hand and the HoloLens has received the interaction event. The HoloLens then processes the color of the button and checks if it is the proper one. In this case, the logic is that checking process in which the next color of the sequence is compared to the introduced one, but any other logic could be implemented here. To finish with this component, it may be possible for the result of the interaction to generate a haptic response (e.g., to change the haptic pattern or to disable the haptic feedback). It is also possible that the device that provides the aforementioned response also delivers relevant information (e.g., the Ultrahaptics device has a Leap Motion on it that generates a representation of the user hand), which can be processed.

- Communication Service: Most haptic devices require a wired connection to a computer, connections that smartphones and wearable AR devices cannot guarantee. Hence, this component will act as a bridge between different clients that collaborate in the process. Besides that, when multiple users are collaborating (e.g., the iOS and HoloLens scenario of Simon), it serves as a networking component. The *Communication Service* has three main parts:
  - A *Network Manager* component is in charge of handling the events on the network.
  - When the event contains updates of the session status, the *Session Status manager* component updates its local representation, which may then be sent to other clients who connect later.
  - Finally, the *Clients Manager* component handles information about client connections (e.g., protocol used, addresses, etc.).

Below is an example of a *session status* in JSON format as it will be stored in the *Communication Service*. As can be seen, the status holds a reference to a root node, from which the remaining nodes can be obtained. In this case, just below the root node, there is an empty node (i.e., it has has an empty mesh associated). Hanging from this node, we have another one (Cube_1), which in this case has an associated mesh but no associated children.

```
{
"rootNode":
{
  "id":"rootnode",
  "isRoot":true,
  "name":"rootnode",
  "transform":{
 "position": { "x": 0, "y": 0, "z": 0 },
 "rotation": { "x": 0, "y": 0, "z": 0 },
"scale":{"x":1,"y":1,"z":1}
},
"children":[
{
"id":"3b2312df-62a0-4488-9383-eaa5bf61596d",
  "isRoot":false,
  "name":"EmptyHolder_0",
  "transform":{
 "position":{ "x": 0, "y": 0, "z": 0 },
 "rotation":{ "x": 0, "y": 0, "z": 0 },
"scale":{"x":1,"y":1,"z":1}
},
  "parent":"rootnode",
  "children":[
{
 "id": "d641b988-b3f3-453f-81b1-7ebfd9767eed","is
  Root":false,
  "name":"Cube_1",
  "transform":{
"position":{"x":-0.037,"y":-0.045,"z":-0.00},
  "rotation":{"x":0,"y":0,"z":0},
"scale":{"x":0.2,"y":0.2,"z":0.24}},
   "parent": "3b2312df-62a0-4488-9383-eaa5bf61596d","c
  hildren":[],
  "childrenIds":[],
  "mesh":{
"vertices":[0.5,-0.5,0.5,-0.5,-0.5,0.5,0.5,0.5,0.5,-0.5,0.5,0.5,0.5,0.5,-0.5,-0.5,0.5,-0.5,0.5,-0.5,-0.5,-0.5,
-0.5,-0.5,0.5,0.5,0.5,-0.5,0.5,0.5,0.5,0.5,-0.5,-0.5,0.5,-0.5,0.5,-0.5,-0.5,0.5,-0.5,0.5,-0.5,-0.5,0.5,-0.5,
-0.5,-0.5,-0.5,-0.5,0.5,-0.5,0.5,0.5,-0.5,0.5,-0.5,-0.5,-0.5,-0.5,0.5,-0.5,-0.5,0.5,0.5,-0.5,0.5,0.5,0.5,0.5,-0.5,0.5],
"normals":[0,0,1,0,0,1,0,0,1,0,0,1,0,1,0,0,1,0,0,0,-1,0,0,-1,0,1,0,0,1,0,0,0,-1,0,0,-1,0,-1,0,0,-1,0,0,-1,0,0,
-1,0,-1,0,0,-1,0,0,-1,0,0,-1,0,0,1,0,0,1,0,0,1,0,0,1,0,0],
  "triangles":[0,2,3,0,3,1,8,4,5,8,5,9,10,6,7,10,7,11,12,13,14,12,14,15,16,17,18,16,18,19,20,21,22,20,22,23]
  }
  }
],
  "childrenIds": [ "d641b988-b3f3-453f-81b1-7ebfd9767eed" ],
  "mesh":{
  "vertices":[],
  "normals":[],
  "triangles":[]
  }
  }
],
  "childrenIds": [ "3b2312df-62a0-4488-9383-eaa5bf61596d" ],
  "mesh":{
  "vertices":[],
  "normals":[],
  "triangles":[]
  }
  }
  }
```

Listing 1: Example of the data model as it is stored in the *communication service*, presented in JSON format.

As said before, the minimum functionality required to augment AR with haptic feedback may be grouped into a single HARP client. However, what happens when all the logic cannot be deployed into a single device? Take for example the Simon prototype in Section 6.2. In this case, there are three different clients: (i) the HoloLens, (ii) the Ultrahaptics, and (iii) the iOS application. The HoloLens client is used to display the AR buttons, and it contains the logic to process the color inputs from the user. The Haptic client receives the 3D model of those buttons, and it generates the haptic representation of them. It is also in charge of detecting the user hand and of sending the appropriate events when it is positioned on the buttons. The iOS Client will also receive the 3D model of the buttons, and it will generate its own AR representation with which the user can interact. Finally, the Communication Service will provide a communication channel common to all clients. The adaptation of these three components to the HARP architecture results are shown in Figure 6.

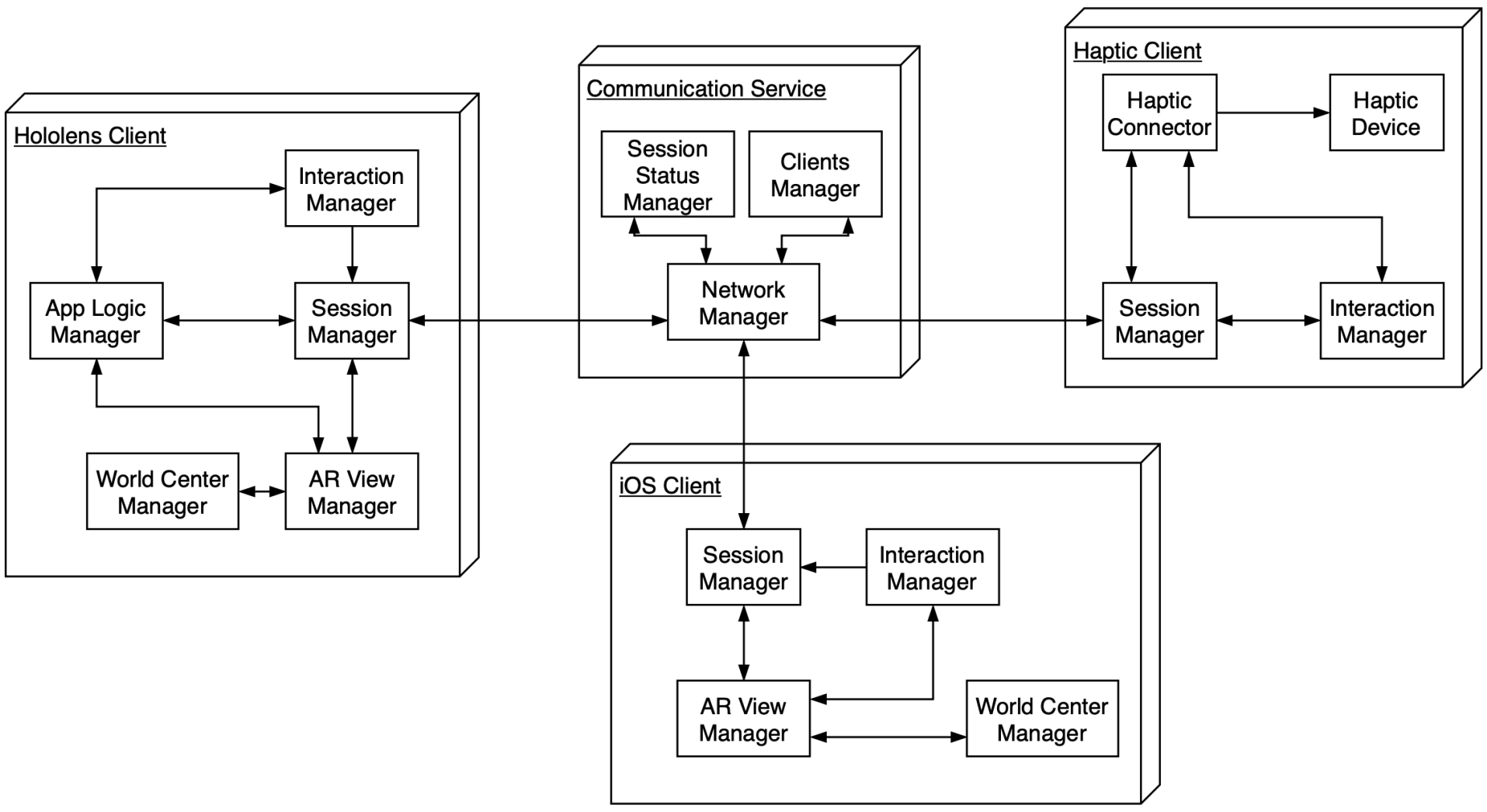

**Figure 6.** Instantiation of the architecture for the Simon prototype with three different clients: HoloLens, Ultrahaptics, and iOS.

As can be seen in Figure 6, the three clients are variations of the one presented in Figure 5. The *App Logic manager*, for example, only appears in the *HoloLens Client*. The information of the pushed buttons is received by the *Session Manager* in the HoloLens, and the game logic reacts to them (it checks if the pushed color is the right one, and if that does not happen, it generates a new color sequence.) The *Haptic Connector* on the *Haptic Client* is in charge of generating the structure required by the *haptic device* to provide the feedback. In this case, the method is the one explained in Section 5.2 (the feature-based representation). In this case, the result is a list of points in the space in which the Ultrahaptics will present a haptic point. The *Interaction Manager* in this client is in charge of receiving hand estimations from the integrated Leap Motion device and of generating the required events when the hand comes in contact with any of the buttons. Its counterpart, the *Interaction Manager* component in the *iOS Client* listens for touches on the screen, and when any AR element has been touched, an event is generated and sent to the *Communication Service*.

## 5. Practical Issues on Haptic Representation

When facing the implementation of a haptic-responsive augmented reality system, there are some specific issues that need to be handled. In this section, we review the solutions given to them when carrying out the implementation for the Ultrahaptics device. Although the presented solutions are technology-specific, they may be inspiring when dealing with alternative technologies. The two main issues are the alignment between the digital, AR content and the haptic feedback, and the representation of the 3D model over the Ultrahaptics (with the consequent apparition of *haptic ghost* *points*). Next, we will address in more detail each one of those problems and we will discuss the solutions adopted for each one of them.

### 5.1. Alignment Method between the Physical and the Virtual World

For the target applications, AR content must be precisely aligned with the haptic feedback. The alignment problem is a constant when combining physical elements and AR content. In short, it consists of establishing a coordinate system common to the digital and the real world. However, the solutions proposed here are tailored to Ultrahaptics and HoloLens but can be adopted too for instrumental systems based on physical vertical feedback.

Two strategies to achieve the alignment are proposed:

- Manual: We have equipped the Ultrahaptics board with three stickers on three of its corners, as Figure 7 shows. In order to perform the alignment, the first task of the user is to move three digital cube anchors using gestures and to place them on top of those stickers. Once this is done, the system is able to calculate a local coordinate system centered on the device. This coordinate system is then used to locate the digital representation of the Ultrahaptics itself on top of which the digital elements will be attached. From this moment on, the haptic representation and the AR content should share the same space. The main problem with this method is that the user may make mistakes when positioning the anchors. Figure 7 illustrates this problem, since the user has positioned some elements on top of the device and some inside it. The more accurate the alignment is, the more precisely the haptic feedback will match the AR content.
- Automatic: In this strategy, the application establishes the coordinate system using mark detection. A QR code or an image is placed next to the Ultrahaptics in a specific position (i.e., known distance and orientation). The device will always be situated in the same position with respect to the mark. By knowing the logical relation between the mark and the device, transformations between both of them are trivial. The framework used to perform the mark detection has been the Vuforia SDK for Unity [49]. Once the system has detected the marker, it shows a transparent red-colored plane over it as will be shown later in Section 6.1. Under certain circumstances (e.g., low illumination or lack of space near the physical device), mark detection may not work as well as expected.

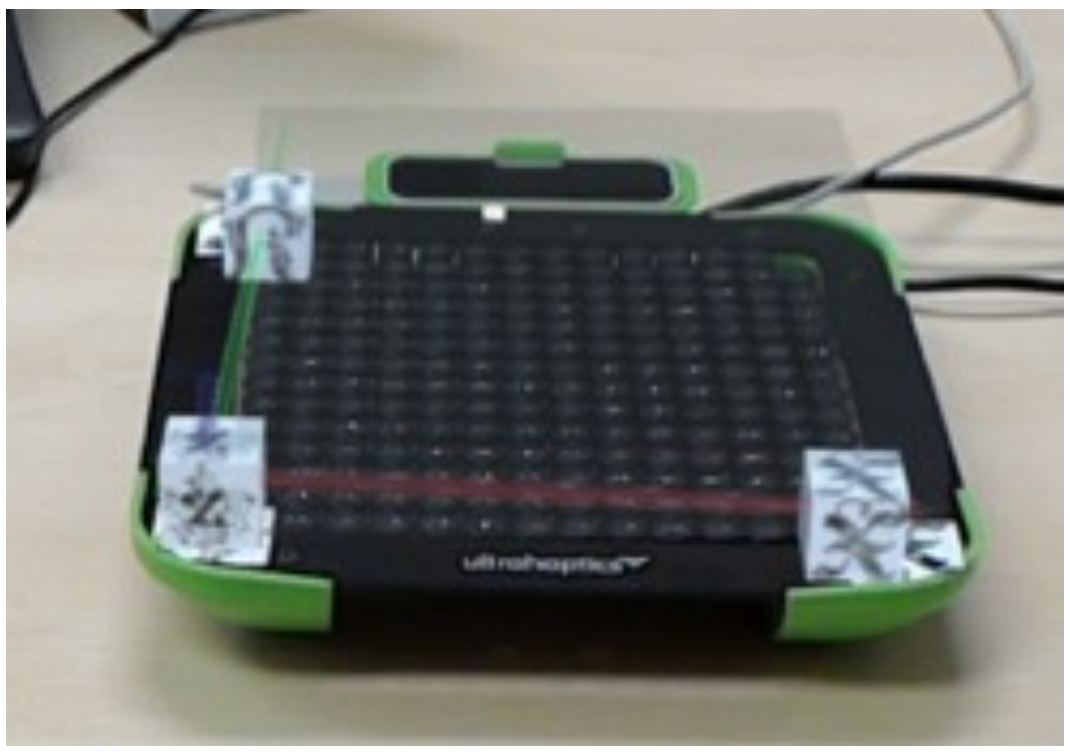

**Figure 7.** Manual alignment over the Ultrahaptics: The user has positioned three digital cube anchors over three physical stickers.

### 5.2. Haptic Representation of a 3D Model in Mid-Air Haptic Scenario

Once the alignment problem has been solved, another key issue is how to "translate" the 3D model (as said before, composed by vertices and faces) into a haptic representation. In Section 3, different approaches to representation considering object fidelity were proposed: high-fidelity and feature-based approaches. We next describe how to implement them in practice, in particular, over the Ultrahaptics. As this device enables the developer to represent *haptic points*, any model to be represented has to be decomposed in a series of haptic point references. In general, the whole process can be divided into the following phases: (a) data model encoding, (b) volume checking (and resizing if required), (c) vertices extraction, (d) duplicated vertices removal, and (e) the haptic representation itself.

The *feature-based representation* is the straightest one to implement, as few reference haptic points are needed (even one may be enough). This could be, for example, the centroid calculated from all the vertices of the model. As has been mentioned, this kind of representation is quite appropriate for very small elements or/and for reinforcing the interactive metaphor for a given representation (e.g., a button).

*High-fidelity representation* may be faced in a different mode depending on whether it is *vertices-based* or *volume-based*. For vertices-based, the procedure is focused on extracting the vertices of the model to represent them on the Ultrahaptics. It is important to note that a volume-adaptation check must be done on the 3D model to guarantee that the representation fits the available volume over the device. It should be mentioned that the received 3D model is encoded as described in Section 4.1, which facilitates the process of extracting the vertices and faces of the 3D model. Then, a vertices redundancy check is done. This is necessary because some platforms introduce redundancies when loading from or exporting to certain 3D formats. From there, all that remains is to translate those vertices into the format required by the Ultrahaptics and to represent them. Nevertheless, the device version used during the development stage (UHDK5) presents a limitation. The number of haptic points that can be presented at the same time is restricted to 4, according to the manufacturer. Thus, at every execution step, a set of haptic points are sequentially represented in a loop. Hence, the bigger the vertices/haptic-point array gets, the worse the representation works. To solve this problem, we introduced padding. Instead of drawing all the vertices in each execution step, we only represent a subset of them in stages. For example, the first one draws with all the vertices in an even position in the array and another draws with only the odd ones.

Two issues has been observed on this procedure. First, the 3D model is only represented through its vertices, so no information was presented between them (there is no edge or face information). Second, as it has been stated, the smaller the number of represented vertices is, the more precise the representation is. In this first iteration, the system “wasted" painting steps by representing points that were not be perceived by the user due to their hand not being at the position to detect them.

If the objective is to perform a volume-based representation, the more detail the representation offers, the more details the user may appreciate. In this case, hand information was introduced into the system using the Leap Motion device integrated on the Ultrahaptics and the generation process suffers certain alterations. In this case, the first step is to divide the 3D model into slices. To do so, we created a logical representation of the 3D shape and encoded it on a 3D matrix. That matrix can be seen as a discretization of the space over the Ultrahaptics, with each cell in that matrix representing a position on the 3D space over the device. For each one of these cells, it is stored as a Boolean value whether it should be represented or not. The more resolution the matrix has, the more details will be perceived by the user. After, the next step is optional. For every triangle of the model, we stored its edges information. If the edge collides with the slice plane, the collision point is marked to be represented. Finally, only the slice in which the user's hand is positioned gets represented. By using this method, the edge information can be easily integrated in the representation and only those points which can be really perceived by the user are shown. Figure 8 shows a schematic vision of how this process works. A red cube appears in the left side of the figure from which the haptic representation will be generated. In the center of the figure appears the representation of the 3D matrix in which only the vertices information has been stored. In the right-most side of the image appears the matrix representation including all the information relative to the triangles of the figure. Each dark cube represents a cell on the matrix, and it will be perceived by the user as a haptic point.

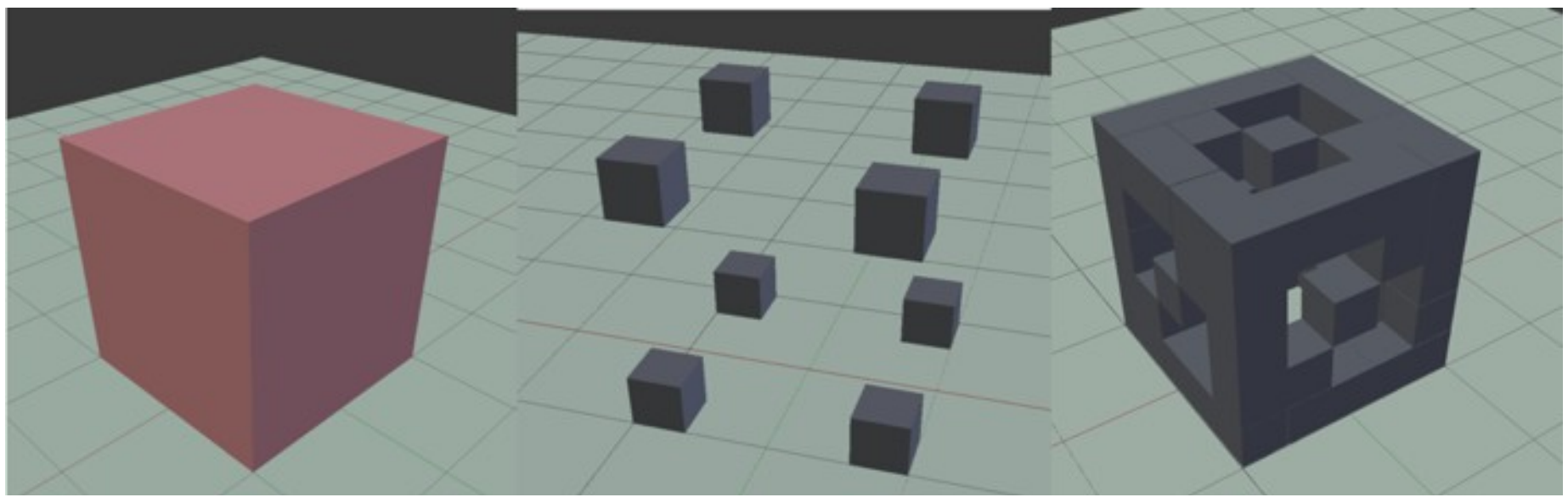

**Figure 8.** Matrix representation of a digital cube: On the left is the original cube. In the middle is the matrix visual representation of the vertices of the cube. On the right side is the same representation but including edge information.

Finally, an issue to remark is that the Ultrahaptics device version (UHDK5, 1.1.0 firmware) used during the development stage presented an additional hindrance regarding the point representation step. This device represents haptic points on top of it that the user can feel with their hand. *Ghost points* appeared when displaying points not centered on the device. Those points were perceived by the user, but they did not represent real information, since they were only a side effect of the use of ultrasound waves to provide the feedback. It is important to note that these points should be ignored by the user, since they do not represent information about the digital element. In principle, there is no solution to eradicate these ghost spots by software means. However, in the tests carried out during the development, we have detected that the users are able to easily ignore these ghost points as long as they have AR support. This may indicate that the AR itself helps when interpreting haptic feedback.

## 6. Development over HARP: Process and Prototypes

To validate the HARP architecture, we have developed two different applications using it. The first one, the "Haptic Inspector", allows the user to haptically inspect a 3D digital model and has been built to analyze the expressiveness and complementarity of the mid-air haptic representation over AR. Functionally, it is built to reproduce shapes and does not include active interaction. The second prototype is an AR-haptic (harpic) version of the "Simon" game [3], and it includes bidirectional exchange of information between the *Ultrahaptics client* and the *HoloLens client*. Both prototypes have been evaluated within two users trials (Section 7). Next, we describe the application implementations over HARP architecture.

### *6.1. Building a New Application: The Haptic Inspector*

In this prototype, a user equipped with HoloLens can haptically inspect a 3D object. By inspecting, we understand to visualize the digital AR 3D model over the UHDK5 and to perceive it in the hand thanks to the haptic feedback. Because we seek to offer a haptic representation as accurate as possible, the *high-fidelity* representation has been used. The application works as follows. First, the user can select a 3D model from an AR menu, which is displayed through HoloLens in front of them. The left side of Figure 9 shows some of the available options in the system. Once a 3D model has been selected, the system generates the proper haptic representation aligned with the AR object over it, as shown in Figure 9. As can be seen in the Figure, a red AR square has been placed over the 2D physical mark to show the automatic alignment process. From here, the user can now inspect the HARptic element or select another one to explore.

Using the Haptic Inspector as an example, the implementation of new applications over HARP is exposed.

The first step is to create and launch an instance of the *communication service* on a computer. Right now, this service has been already programmed, and it contains the basic logic required to manage the sessions. Network management operations inside the *Communication Service* have also been implemented. Hence, no additional coding is required in this step. The final result is a Node.js [50] based application which allows receiving connections from different technologies and is capable of separating the network layer from the logic one.

The second step is to create the *(HoloLens) client*. This platform works on Unity and the MRTK SDK for Unity. We have generated a component that, on this platform, is known as a *prefab*: a combination of scene elements and scripts ready to be imported into any project. This prefab automatically generates an instance of the *session manager* component containing all the required functionalities to establish communication with the *communication service*. The *AR view manager* component is, in this case, transparent to the programmer. By using the MRTK, any element added to the Unity scene will be shown by the HoloLens. To generate the *world center manager* for the Ultrahaptics, any of the alignment strategies identified in Section 5.1 can be used. However, both approaches have been programmed with the Ultrahaptics scenario in mind. In case it is necessary to establish the alignment for an alternative device (which, for example, is not easily located on a table), it will be necessary to specifically code the alignment procedure. In any case, the minimum functionality that should be provided to the user would be to position the center of the session anywhere in the real world. Next, the *app logic manager* is implemented. In this case, whenever the user selects a 3D model from a menu, that model is shown over the Ultrahaptics, added to the session status, and sent to the *communication service*. To finish with the *HoloLens client*, the last step is to design the *interaction manager* component. This module is quite straight to implement. MRTK SDK itself provides a way for detecting when some gestures have been done by the user (e.g., tapping a 3D object, which would be the equivalent to clicking it). For each element in the Unity scene with enabled interaction, the corresponding method must be implemented.

The third step consists of implementing the *Haptic client* using the Ultrahaptics and Leap Motion Unity libraries. Since this client is also based on Unity, we are able to import the same *session manager* prefab as described before. Within this prototype, the *interaction manager* component is only in charge of detecting the user hand and of providing its position to the *haptic connector*. Finally, in order to create the *haptic connector*, we had to implement the logic for haptic representation as described in Section 5.2. This logic is already available in HARP, although it can be exchanged or extended. In this case, as the result of this component and based on the user's hand position, a list of haptic points is generated and sent to the *haptic device* to be represented.

In addition to the predefined set of geometric figures, we have enabled the introduction of arbitrary 3D models in the system. However, due to the low resolution that Ultrahaptics allows, the details of these models fade away.

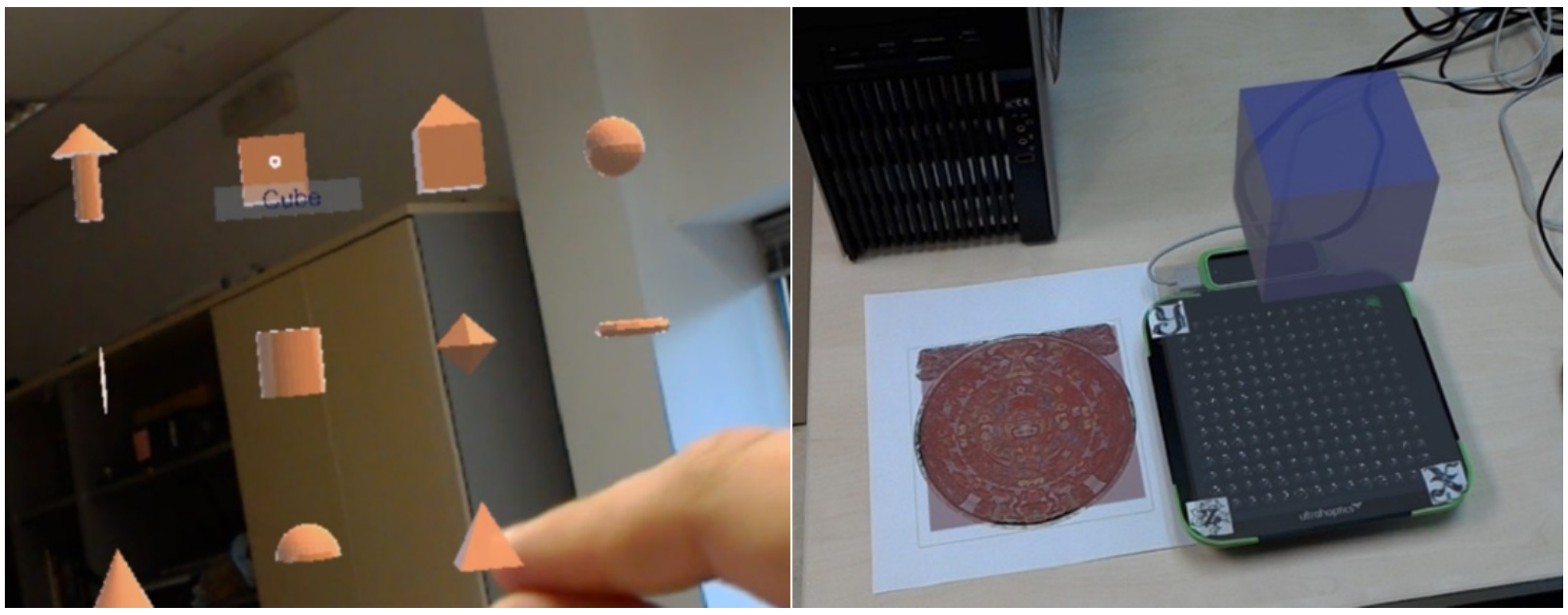

**Figure 9.** First proof of concept: Haptically inspecting a digital object. On the left side appears all the possible primitives that the application offers. On the right side is the cube primitive ready to be inspected by the user.

### 6.2. Building Haptically Interactive Applications: The Example of the HARptic Simon Game

Four AR buttons, all with different colors, are located over the Ultrahaptics in order to mimic the physical game created by Baer and Morrison (1978). In our implementation, the system generates a sequence of colors that the user must repeat by pressing the AR buttons with their hand. The objective is to replicate the sequence, which will grow over time, without failing. The participant will play the game through either HoloLens or by using a mobile iOS client. A multiplayer version of the game can be enabled through the use of both devices. Figure 10 shows the color buttons over the Ultrahaptics, from the point of view of the HoloLens.

The main features of this proof of concept are the following:

- The haptic representation of the 3D buttons is generated through their centroids, using the *feature-based* representation. This is possible because buttons are small and haptics are used to enhance their perception, i.e., to locate them and to tangibly show the resistance that a non-pressed, real button naturally shows.
- The position of the user's hand is tracked with the Leap Motion integrated in the Ultrahaptics. This allows us to make the buttons react to the position of the hand. As the user presses a button, it will get closer to the device, and to simulate this resistance, the ultrasound intensity will rise. When the user effectively presses the button, it generates an event that includes the color information. Once the user releases it, the button "returns" to its original position, with the ultrasound intensity back to the normal level. In this case, the status of the session is updated from the *haptic client* and changes are notified to HoloLens.
- When the user pushes a button, an *interaction event* is generated. This event is just a message that encapsulates certain information about the user having interacted with a digital element (e.g., an object that generates the event, time, and session to which it is attached). This event is then sent to the *communication service* and forwarded to all the devices that are participating. This kind of event is not exclusive of the *haptic client*, as they could be generated anywhere in the architecture.

  Let us suppose that the user has pressed the blue button. The *haptic client* encodes that information within an *interaction event*, which is sent to the *HoloLens client*. When the HoloLens receives that message, it decodes to which element of the scene it refers (each node has an unique ID), checking if any reaction is expected. If so, it checks whether the next element of the sequence that the user must enter is the blue color. If that is correct, the user is one step closer to matching the sequence. Otherwise, there has been an error, so the system alerts the player and shows them the same or a different sequence. In summary, this kind of event allows great flexibility when integrating different devices in the architecture ecosystem.
- The logic of the whole application is located on the *HoloLens client*. The *haptic client* is able to receive the 3D elements and to generate proper haptic representation without obstructing the logic of the application. This feature is highly desired, since at any time it is possible to replace the Ultrahaptics with any other haptic device.
- A second participant using the mobile *iOS client* can receive the status of the session as seen by the HoloLens user. Going one step further, this user can interact with the buttons by touching them on the device's screen. This interaction is integrated in a seamless way with the rest of the components.
- For creating the *iOS client*, we followed the same steps explained in Section 6.1 but, this time, adapted them to the mobile system. As a result, we can offer a library similar to the prefabs mentioned in Section 6.1 but for the iOS platform.

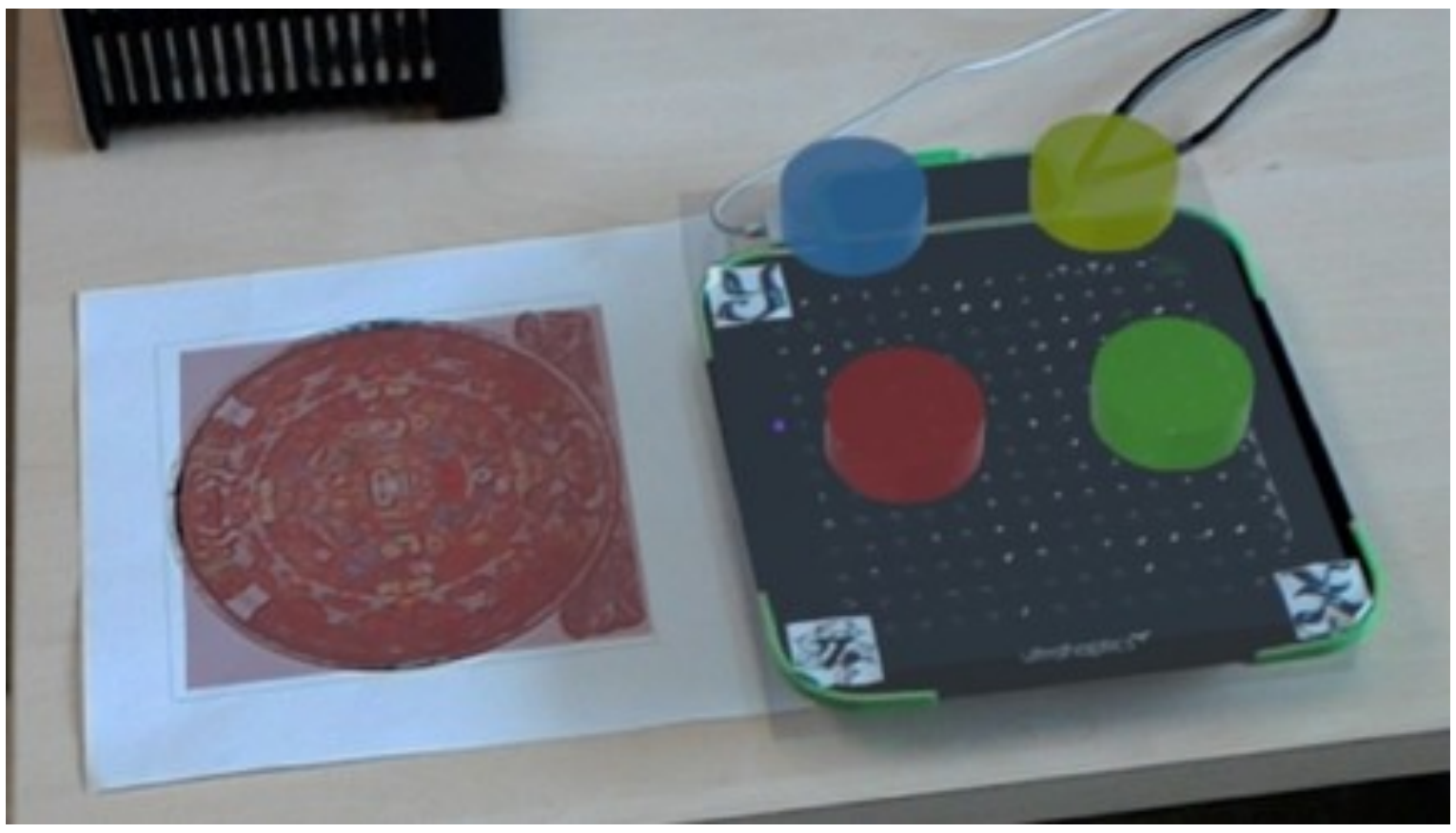

**Figure 10.** Second proof of concept: Simon game based on the combination of Augmented Reality (AR) and haptic feedback.

## 7. Validation with Users

### 7.1. User Study Objectives

In this section, we describe the trials carried out over the two prototypes described in Section 6. The purpose of these tests is twofold. On the one hand, we seek to technically validate the prototypes built over HARP, with real users to demonstrate that the architecture enables to deliver haptically augmented AR applications, both to increase perception (Haptic Inspector) and to improve interaction (Simon). On the other hand, we seek to better understand the contribution of haptic feedback to AR-based tasks, both in functional and acceptance terms. First, we explore the perception of the contribution given by haptic feedback with AR visualization for inspection tasks. Second, we explore the expressiveness of Ultrahaptic-based haptics for shape identification procedures. Third, we study the usefulness of haptic aids for AR resizing tasks. Fourth, we explore haptic adaptations for interaction metaphors with mid-air haptics and AR visualization. Finally, we study the grade of acceptance of AR with mid-air haptics with respect to a mobile baseline. The details of user study are presented below.

### 7.2. Participants and Apparatus

The study was carried out in July 2019 with a total of 9 independent volunteers recruited at the university (PhD students, professors, and master students). None of them had any previous knowledge about the project or the technology in use. Their ages range between 23 and 57 ($\mu$: 32.62, $\sigma$: 12.17); all of them were males: 44.4% of users use AR applications sporadically and 11.1% use it once a month, while the remaining 44.4% never use AR. The tests have an average time of 62 min ($\sigma$: 10.48). The apparatus in use for the tests are the two prototypes described in Section 6. In the *Haptic Inspector* case, a slight modification of the prototype has been made to prevent the user from selecting the 3D figure to be inspected; the system will do it automatically for them by following a predetermined sequence, as explained in the next Section.

### 7.3. Procedure, Tasks and Experimental Design

Each user has been asked to fill in an informed consent as well as a small survey regarding their previous experience with AR. During the test, the facilitator used a written script, so all participants received the same information. The study has been divided into 5 tasks, labelled from T1 to T5. Tasks 1 to 3 make use of the Haptic Inspector prototype of Section 6.1. Tasks 4 and 5 are based on the Simon prototype of Section 6.2. In order to avoid the learning effect due to specific use of Ultrahaptics, before starting the test, each user experienced a short application example (moving bubbles, extracted from the Ultrahaptics Demo Suite). In this short demo, participants position their hand on the Ultrahaptics, so they can perceive the feedback and the interaction range. The facilitator has also been in charge of aligning the systems through the automatic mode described in Section 5.1.

Next, we detail the purpose of each task together with its workflow:

- **Task 1, perception of the contribution of haptic feedback over AR visualization for inspection tasks**: This task has been designed to offer the user a first contact with the combination of haptic feedback and AR or with HARptic. The basics of the system are introduced to the participants: they will be able to haptically inspect different geometrical figures that will be visible through HoloLens, and they can *inspect* with their hand. Participants are not told how to actually inspect the figures; their own interpretation of the action will condition the way they interact with the views. This process is repeated for the 10 images in Figure 11. For each figure, the exploration time is saved. Additionally, the user is asked to rate from 1 (very difficult) to 5 (very easy) to which extent the figure would be guessable only through the delivered haptics (no visualization available). The images chosen to be included in the test (for Tasks 1–3), are divided into the following groups: A first group of basic geometries (*pyramid, cone, sphere, hemisphere, cube, and cylinder*), a second one of more complex geometries (*octahedron and torus*), and one last group of compound figures (*house and arrow*) thought to simulate objects. Figure 11 shows how figures are presented to the users.

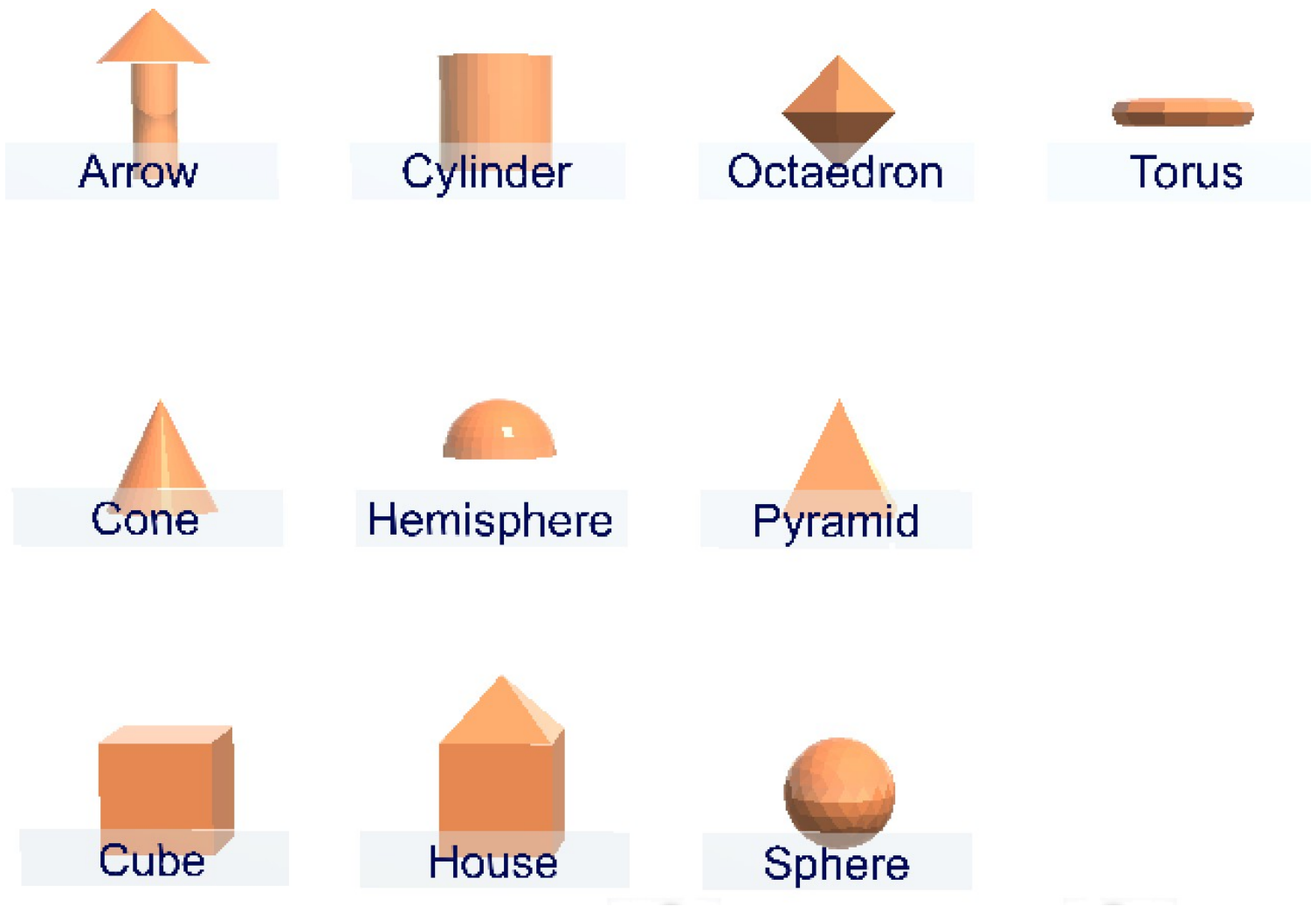

**Figure 11.** Set of figures used for the development of Tasks 1–3.

- **Task 2, Ultrahaptics-based haptics expressiveness for shape identification**: In this task, the user has to guess the shape of the object under exploration only relying on haptic feedback. The system provides the haptic representation of different 3D figures, and the user has to guess the object identity from the ones available in a list. The participant's guesses are stored as well as the time required to give an answer. The number of hits is not revealed to the user until the entire task has been completed, since after the task, a questionnaire on their perception of the task difficulty is administered. The figures are the same as that in Task 1 without repetitions. Figure 12 shows three participants while they are exploring the harptic figures. The captures have been taken using the HoloLens client.

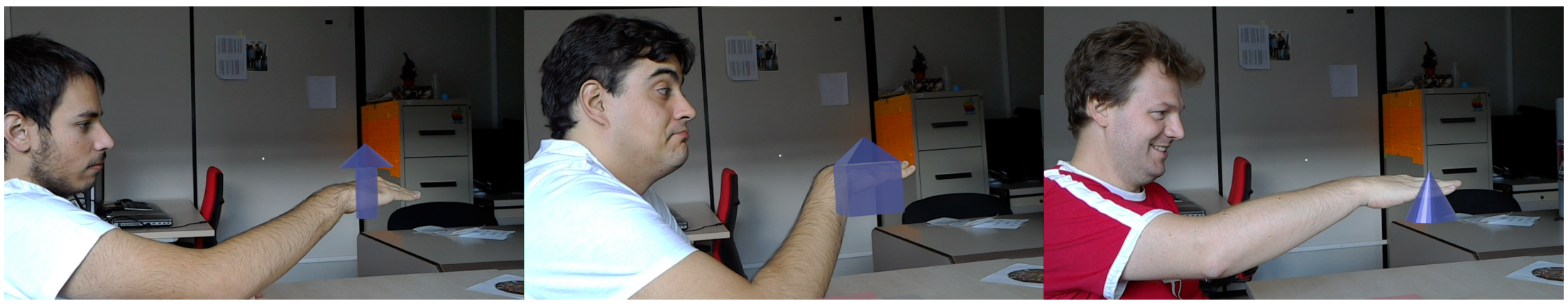

**Figure 12.** Three participants trying to guess a figure from haptic feedback only in Task 2.

- **Task 3, usefulness of haptic aids for AR resizing tasks**: For this task, visual and haptic representations are again merged and delivered. In this task, the system presents the participant a certain figure with preset dimensions. We have preset a virtual ground at 17 cm from the surface of the Ultrahaptics; thus, every figure stands over this ground. With this information in mind, participants are asked to resize the figure to reach a given dimension (height). A haptic anchor on the target height is deliver through the haptic device. To perform the task, the participant must put their hand over the Ultrahaptics and move it up and down as required. The process should be supported by estimations using the augmented reality. Once a participant determines that the target height has been reached, the actual size of the model is recorded as well as the required time to perform the process.
- **Task 4, mimicking real-life interaction metaphors for AR with mid-air haptics**: Both, this task and Task 5 are based on the Simon prototype. Again, the user is wearing the HoloLens which shows four colored buttons over the Ultrahaptics. Once the task starts, the user has two and a half minutes to complete as many sequences as possible. The participants are told to press the buttons, with no reference about how to do it. If the user hits the sequence, the system generates a new one with one more color. Otherwise, if the user fails, the system generates a new sequence with the same size but with different colors and the fail is saved. Besides that, the number of correct sequences is stored.
- **Task 5, acceptance of AR with mid-air haptics with respect to a standard mobile baseline**: At this stage of the test, the participant is already familiar with the system. In this task, the user is asked to participate in a collaborative Simon game in two rounds. In the first round, the user will be interacting with the game through HoloLens and UHK while the facilitator will be using an AR iOS version of the game. The goal is the same as in Task 4, i.e., to hit as many correct sequences as possible within two and a half minutes. However, in this case, the user and the *facilitator* must act in turns. First, the HoloLens client has to push a button and, then, the iOS client has to push the next one and so on. Both users receive the sequence at the same time, each one in their own device; at any time, it is possible to ask for help or instructions from the other user. Once this is finished, both users exchange the positions. Now, the *facilitator* works with the HoloLens and the Ultrahaptics and the participant uses the iOS application. A new round begins then, just like the previous one.

Figure 13 summarizes the workflows for the test and the questionnaires to be filled in at each stage.

The experiment was designed as a within-subject one, with different factors depending on the task. For Task 1 and 2, factors were the availability of AR (two conditions: yes/no) and 3D figures (10 conditions). We have a total of 180 trials (10 figures × 9 users × 2 AR condition values (on/off)), with Task 1 including AR visualization and Task 2 relying only on haptics for guessing. As dependent variables, the execution time, interpretability values, and hits were measured. For Task 3, the factors are the 3D figure shape (5 conditions). Hence, a total of 45 trials have been gathered (9 users × 5 figures, each one with the desired height associated). In this case, distance to target height and time have been the dependant variables. Regarding Task 4 (factor: haptic availability, 2 conditions, on/off) and Task 5 (factor: technology in use, two conditions mobile iOS vs. HoloLens + Ultrahaptics), the dependent variables gathered have been the number of correct sequences by the user as well as the number of fails together with user preference.

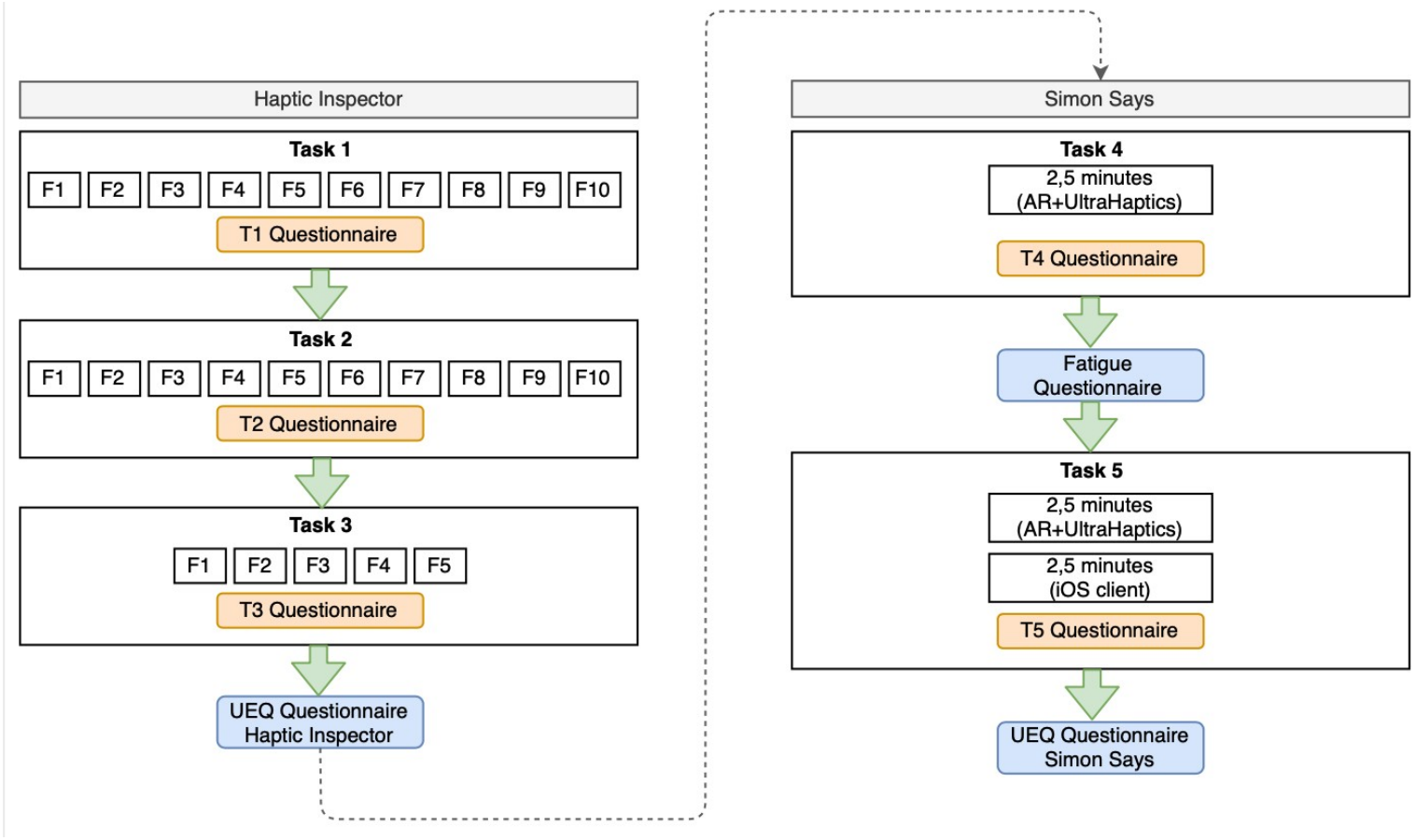

**Figure 13.** Stages of the users study.

## 8. Experimental Results

Following, we summarize some of the study results.

### 8.1. Perception of the Contribution of Haptic Feedback over AR Visualization

In Task 1, as explained in Section 7.3, participants were told that they could *inspect* the figures with their hand. Starting from this premise, users can be split into two groups taking into consideration their exploration movements. Five participants moved their hand up and down to explore the AR figure, interpreting the slices that were generated in the palm of the hand. This group was coincident with the design hypothesis and the implementation described in Section 5.2. On the other hand, 4 participants tended to explore the figure following the edges of its volume. For example, a participant tried to explore the sphere by moving the hand over its entire surface. Due to the operation mode of the Ultrahaptics, if the ultrasound is blocked with part of the hand, the haptic feedback is not properly received. Hence, in this case, it is not possible to inspect the bottom side of the figures with the palm facing up.

When asked if they think that, when provided haptic feedback, high-fidelity volume-based interpretation is sufficient by its own to identify the figures, aggregated results reveal an average rating of 2.11 over 5 ($\sigma$: 1.19), with 1 being completely insufficient and 5 being completely sufficient. Thus, according to the participants, haptic feedback does not seem sufficient and AR visual info is needed.

Table 1 collects the opinion of the users with respect to the hardest and easiest figures to interpret based only on haptic representation. Seven participants identify the torus as the most complicated figure to interpret, while five think that the cube and the cone are the easiest ones. The most contradictory shape is the arrow, with a coincident number of participants (4) voting as it being hard or easy.

**Table 1.** Number of votes for the hardest (left) and easiest (right) figures as voted by the participants.

| Hardest Figure | (Number of Votes) | Easiest Figure | (Number of Votes) |
|---|---|---|---|
| Torus | 7 | Cube | 6 |
| Arrow | 4 | Cone | 5 |
| Sphere | 4 | Arrow | 4 |
| House | 3 | Sphere | 3 |
| Cone | 2 | Pyramid | 3 |
| Cube | 1 | House | 2 |
| Pyramid | 1 | Cylinder | 1 |
| Cylinder | 1 | Hemisphere | 1 |
| Hemisphere | 1 | Octahedron | 1 |
| Octahedron | 0 | Torus | 0 |

During the tests, several participants have indicated that features such as the tip of the cone or the tip of the pyramid and the octahedron are much better appreciated. At the time of guessing in Task 2, these users have tried to find those characteristics that seem relevant to them.

Figure 14 shows a box plot of the interpretability value given by the users to each figure. In the image, median values for each figure are represented as circles. As it can be seen in the image, the figures that present a greater disparity of results are the hemisphere and the torus. On the contrary, figures in which the users have most agreed are the pyramid and the cone.

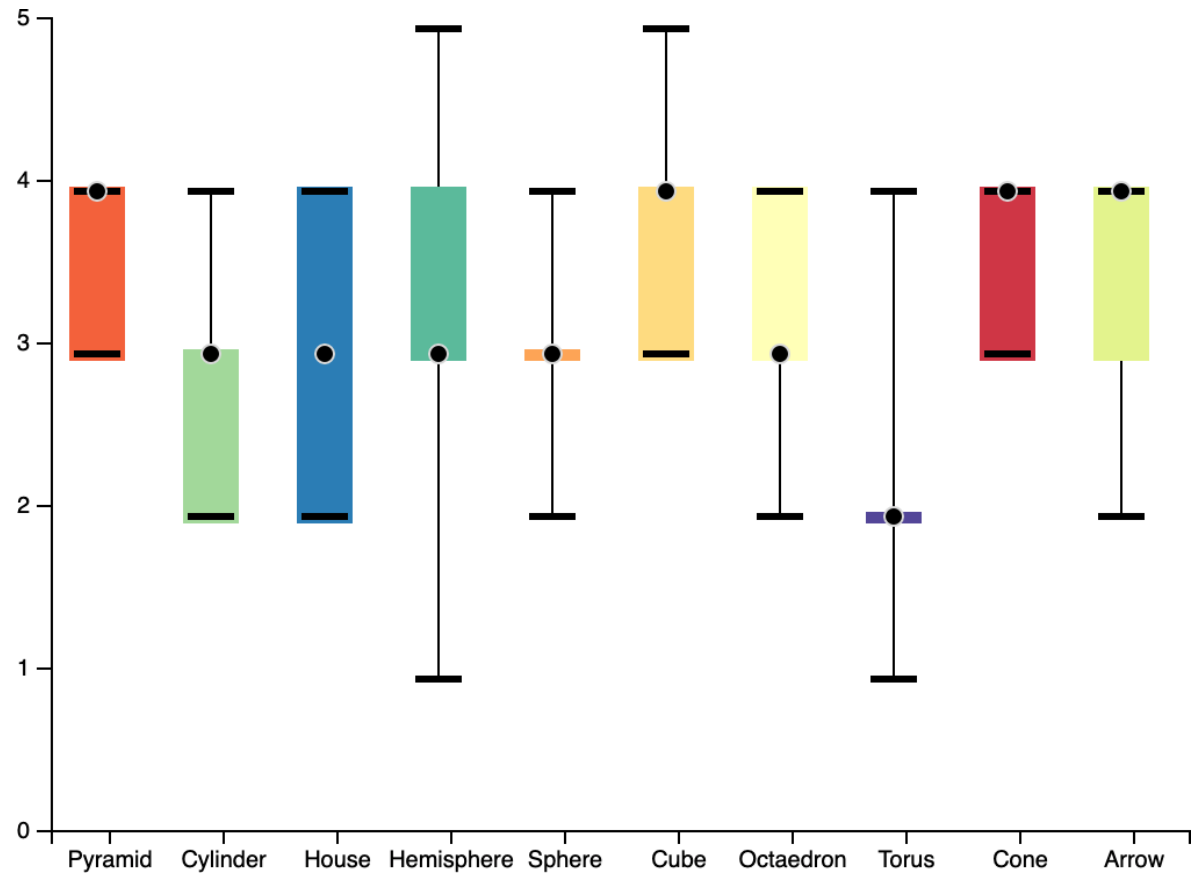

**Figure 14.** Box plot of the interpretability given to each figure by the users, with 1 meaning that the figure is not interpretable and 5 meaning the figure is perfectly interpretable.

Figure 15 gathers the times required by each user on Task 1 since they begin to explore the figures until they form an opinion about it. The light blue line shows the average times for each figure. Times required by users range between 5 and 65 s. There is a statistically significant difference between inspection times among users, as revealed by a one-way ANOVA test ($F(8, 81) = 5.627$, p«). A Tukey post hoc test shows that the inspection time was significantly lower for U1, U2, U5, and U6 when compared to U7 and U8.

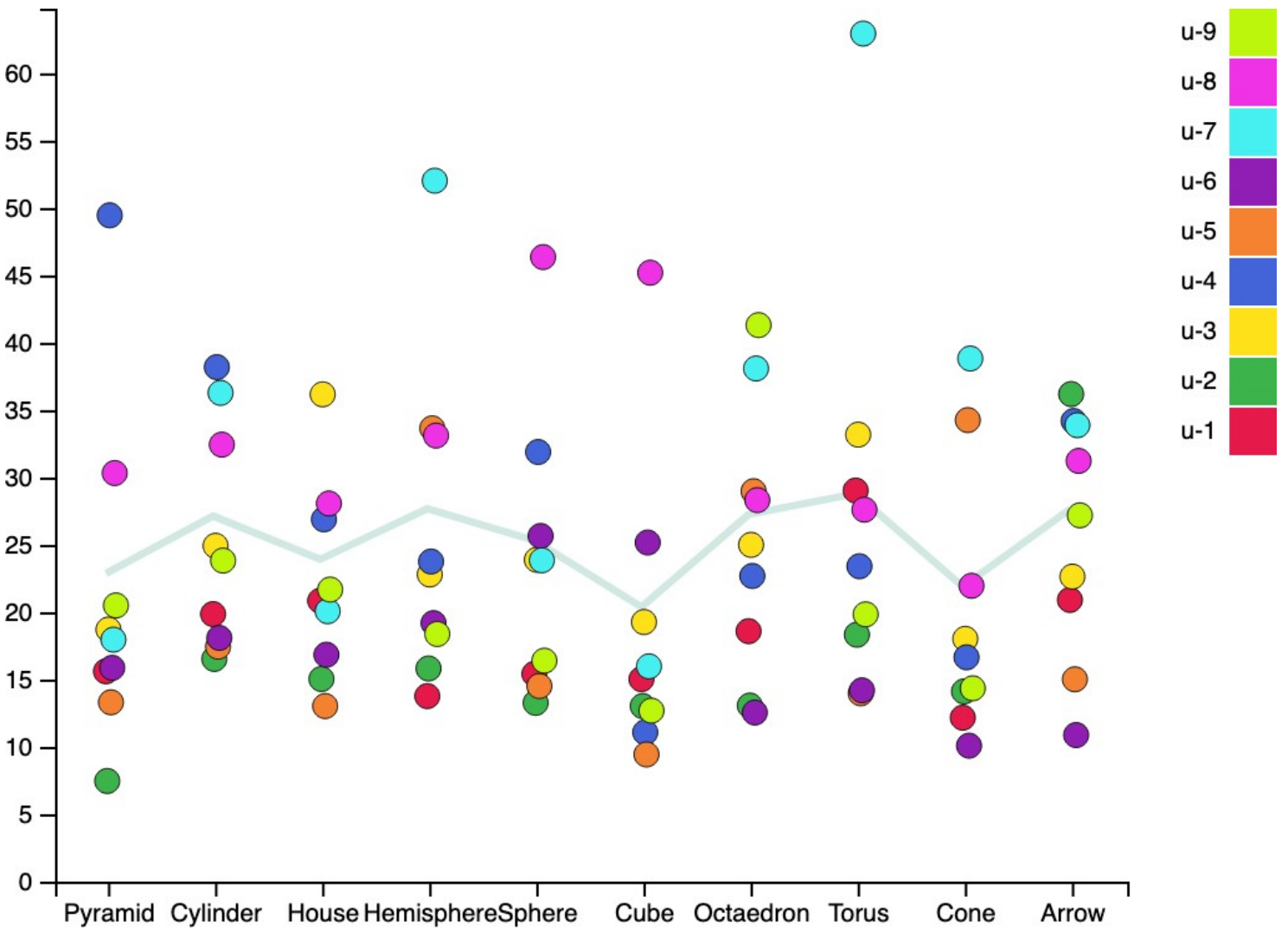

**Figure 15.** Time required by each user for the inspection of the figures in Task 1.

*8.2. UHK-Based Haptic Expressiveness for Shape/Volume Identification*

In Task 2 (guessing the figure only with haptic feedback support), the expressiveness of the mid-air haptic representation for different geometric figures is analyzed: 44.4% of users indicated that the process of guessing figures seems extremely difficult (1 over 5), and 44.4% found it difficult (2 over 5). Only 1 user found it easy. Recall that the user answers the relevant questionnaire without getting any confirmation on the hits and failures. Despite its limitations, results indicate that participants are able to interpret more figures than expected, offering an acceptable success rate. Times required by each user to guess the shapes can be seen in Figure 16. A Wilconxon Soigned-Rank Test indicates that the difference between inspection and guessing times is not statistically significant ($Z = -1.630$, $p = 0.103$).

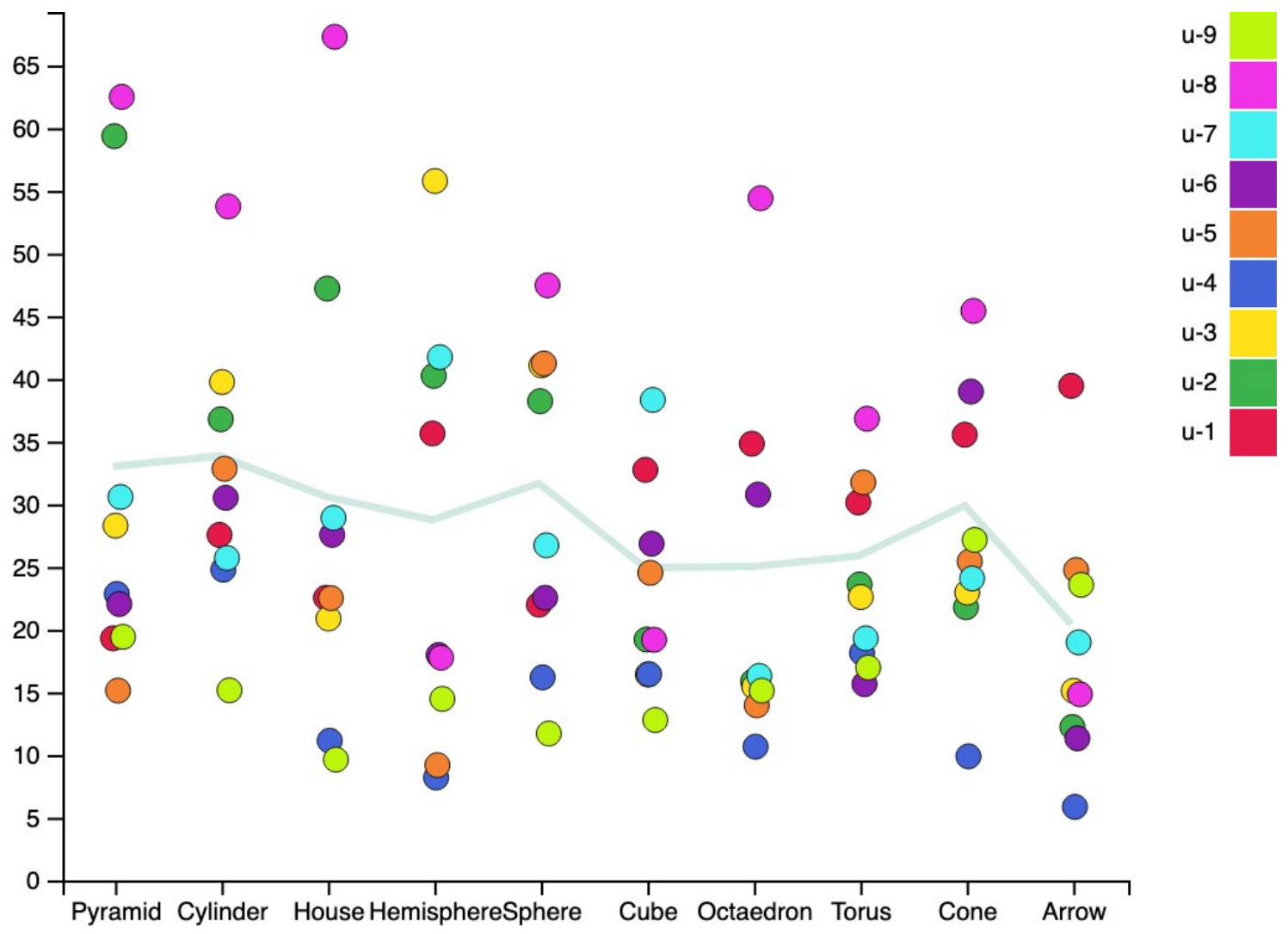

**Figure 16.** Time required by each user for the guessing process for each figure in Task 2.

Figure 17 presents the confusion matrix between the correct answers and the participants' guesses in the form of a heat map. In the vertical axis appears the correct answer, and in the horizontal one appears the user guess. On the diagonal of the matrix, the number of right answers for each figure can be seen . The more green a cell is, the more users guessed correctly that figure. As can be seen in Figure 17, the *cone* has been guessed properly 55.5% (5/9) of the time, while the *torus* has the most hits with 77% (7/9). The *house* on its side has been properly guessed the 66.6% (6/9) of the time. Although users have indicated that haptic feedback is not enough to identify figures, the data indicates that this is possible. In fact, by choosing a set of easily recognizable and non-misleading figures (e.g., cone and pyramid), it would be possible to transmit graphic information to the user in a haptic manner. The presented matrix serves as a guide to select that hypothetical set of figures.

Most users indicated that, when identifying the *torus*, they were not able to properly distinguish the central hole. Instead, they did the divination on the basis that this figure had the lowest height of all. This problem is due to the lack of accuracy of the device as indicated in Section 4.

Figure 18 shows the performance for each participant and the relationship between the figures perceived by users as more difficult (black lower triangle, labeled as 1/5) and easier (blue lower triangle (5/5)) and their actual results (upper triangle in green is a hit, while red is a fail). It would be expected that most hits have occurred over figures labeled as easy, with failures concentrating on those labeled as difficult to interpret. Taking for example the participant U7, we can see that the three figures that the user found easy to interpret (hemisphere, house, and pyramid), were not afterwards guessed. However, two of the figures that the participant said were hard to interpret were two hits (cone and octahedron). In general, it seems that users' perception of figures that are easy and difficult to interpret does not correspond to the results obtained. However, it seems that there is a set of figures that are the most successful in general (cone, house, and torus). We have grouped participants into two groups y taking into consideration the number of hits (G1: ≥5 hits; G2: <5 hits). A Shapiro test is carried out over the hit frequency, and the test cannot reject the null hypothesis of normality for the distribution ($W$ = 0.94, $p$ = 0.61). Then, a Welch Two Sample test for unequal variances cannot state a real difference in performance between the two groups.

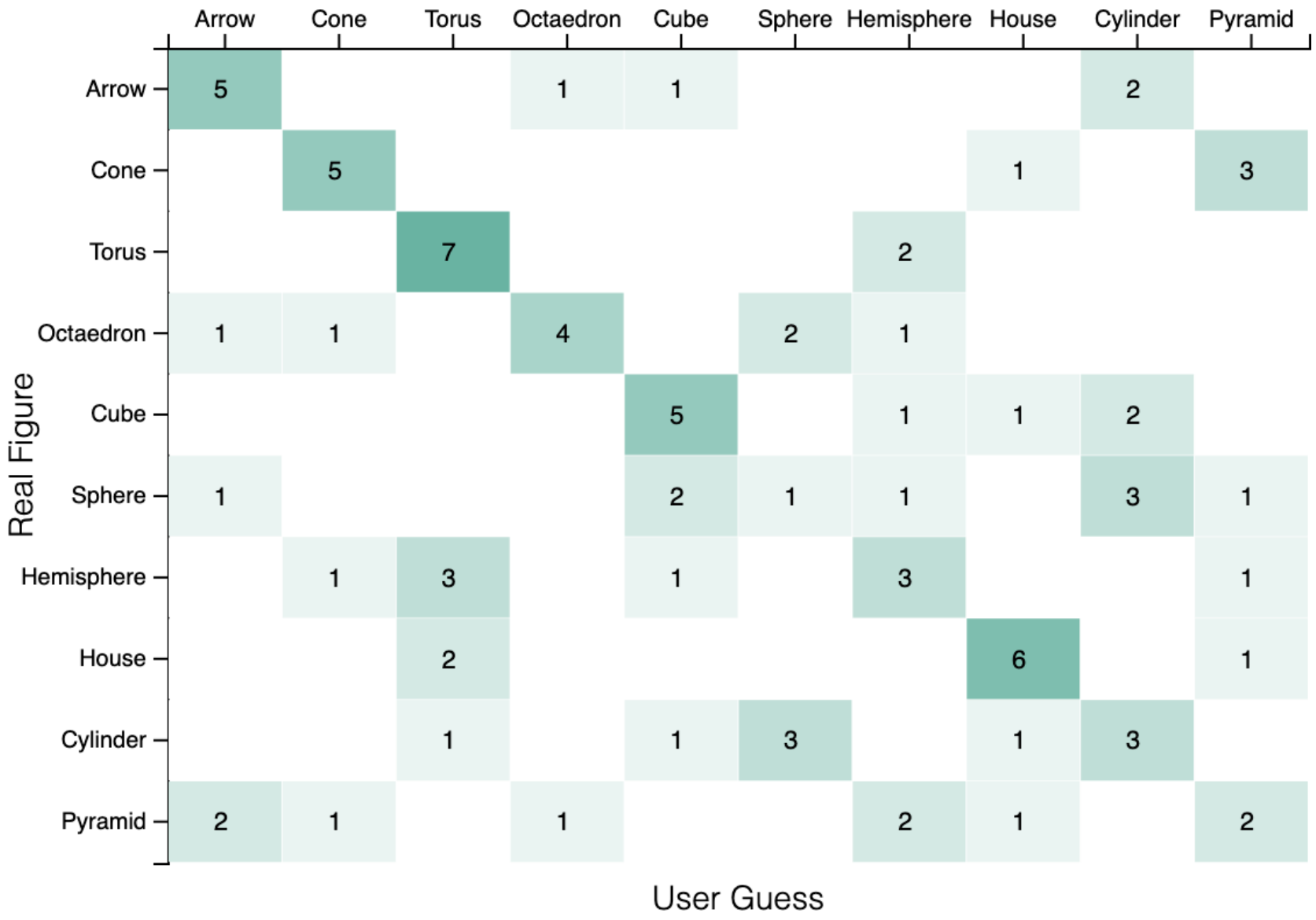

**Figure 17.** Confusion matrix heat map of the Task 2 guessing process.

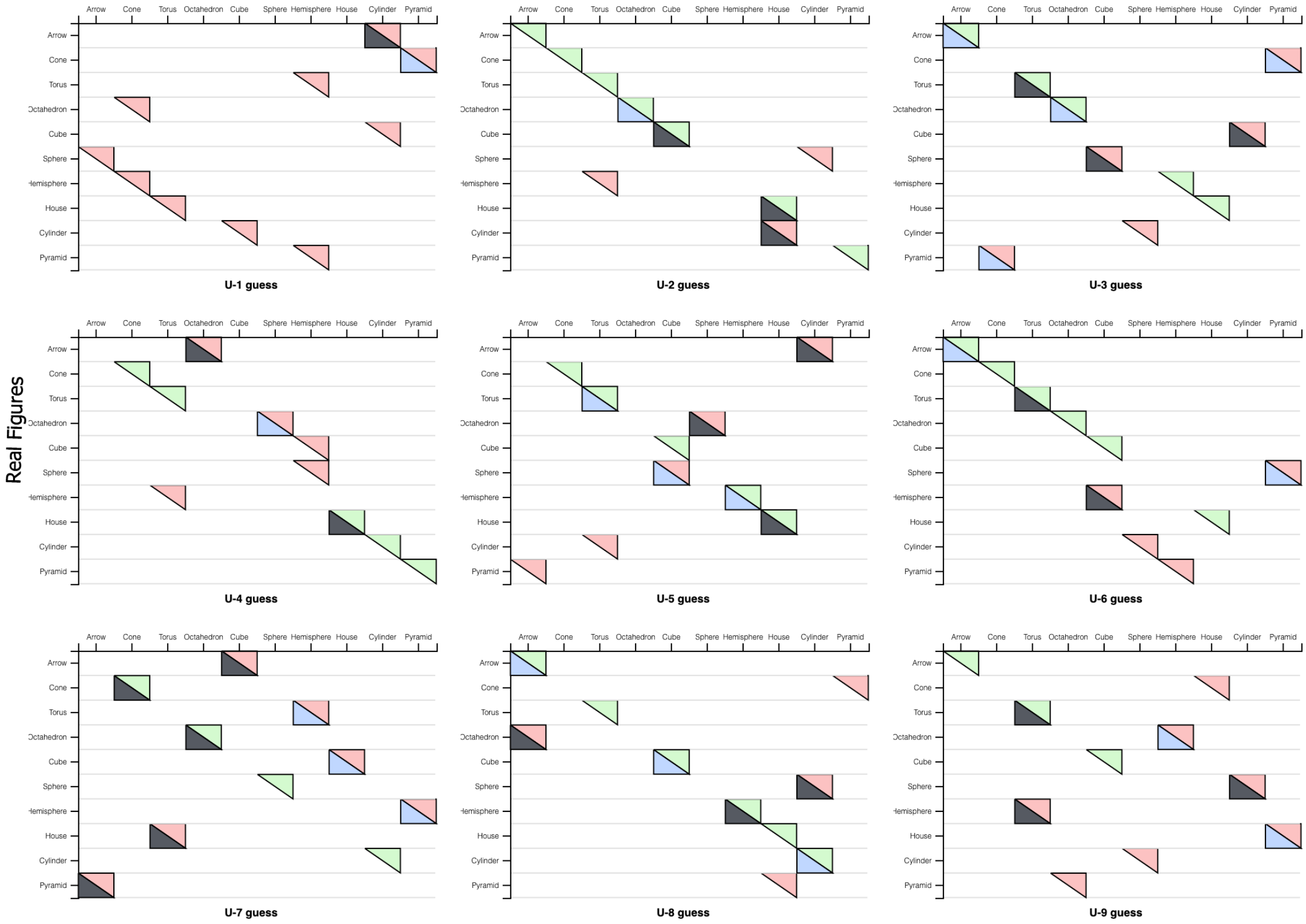

**Figure 18.** Relationship between the perception of users about the difficulty of the figures and their result when guessing. The result of the divination of each figure is represented for each user (Task 2). Each attempt generates a rectangle divided into two triangles. The upper triangle shows if the user was right (green) or failed (red) in their attempt to guess. In the lower triangle appears the user's perception about the difficulty of haptically identifying the figure (Task 1), especially difficult (black color) or easy (blue color).

For generating the haptic representation of the 3D models, the process indicated in Section 3 has been used. At the moment of truth, the users looked for relevant characteristics that, more than helping them to identify the figures, would help them to differentiate those figures from others. As a final note on Task 2, if haptic representation of the figures is improved by emphasizing the key points and by removing irrelevant information, in our opinion, it would be possible to improve the whole guessing process.

### 8.3. Usefulness of Haptic Aids for AR Resizing Tasks

In Task 3, users are proposed to resize some of the figures above (octahedron, cylinder, pyramid, cube, and sphere) with the help of a haptic reference that is of the size to be reached. From there, they have labeled the difficulty of manipulation process with a 3.57 on average ($\sigma$ :1.04) and a 3.71 ($\sigma$ : 1.03) regarding how intuitive the process is (1 is not intuitive at all, and 5 is very intuitive). When rating whether the haptic feedback helped them for the manipulation process, there seems to be a discrepancy (2.12 on average, $\sigma$ : 1.45). The fact is that some users ignored the part of the task statement in which they were told that the system helped them to identify the right position by means of haptic feedback. Therefore, when asked if they would get rid of haptic feedback and rely only on visual AR estimations, 57.1% of the users said yes, 28.6% answered no, and the remaining 14.3% said it was not clear for them. Among some reasons in favor of haptic feedback, they stated that *"it is easy and intuitive"*, and although they are able to only estimate with AR clues, haptic feedback was necessary to have greater precision. One user suggested that, instead of setting the haptic feedback at the right, fixed position, it could be better to generate different haptic intensities depending on the height. That is, the closer the hand gets to the correct position, the greater the intensity of the feedback gets. Table 2 shows the offsets between the target heights in the resizing process and the final values participants produced, measured in centimeters. As can be seen in the table, participant U5 had the best results while U4 and U7 did the worst. These last two are part of the group that ignored the part of the statement about haptic feedback. Actually, a one-way ANOVA test shows that there is a statistically significant difference with respect to the accuracy achieved among the participants ($F(8, 36) = 6.366$, p«), and a Tukey post hoc test reveals that U1, U4, and U7 conform to an homogeneous group with lower performance. The same experiment applied to the time for task completion does not show a statistically significant difference between groups ($F(8, 36) = 0.656, p = 0.725$). Therefore, in a relatively complex task such as estimating height based on AR, the introduction of haptic feedback has allowed most users to perform more than acceptable, with minimal errors.

**Table 2.** Offset in centimeters between the required size and the one entered by the user: The order of the figures is random.

| | Error in cms | | | | | | |
|---|---|---|---|---|---|---|---|
| | **F1** | **F2** | **F3** | **F4** | **F5** | **Avg** | **Std. Dev.** |
| U1 | 3 | 3 | 1 | 1 | 1 | 1.8 | 1.09 |
| U2 | 0 | 2 | 0 | 0 | 0 | 0.4 | 0.89 |
| U3 | 0 | 0 | 1 | 1 | 0 | 0.4 | 0.54 |
| U4 | −1 | 1 | 2 | 3 | 5 | 2 | 2.23 |
| U5 | −1 | 0 | 0 | 0 | 0 | −0.2 | 0.44 |
| U6 | 1 | 1 | 0 | 1 | 0 | 0.6 | 0.54 |
| U7 | 3 | 6 | 2 | 5 | 4 | 4 | 1.58 |
| U8 | 0 | 1 | 1 | 0 | 0 | 0.4 | 0.54 |
| U9 | −1 | 2 | 1 | −1 | 0 | 0.4 | 1.30 |

### 8.4. Mimicking Real-Life Interaction Metaphors for AR with Mid-Air Haptics

In this Section, the results of Task 4 and the Simon prototype are presented. The operating mode of the haptic buttons is as follows. The haptic point representing a button is suspended at a certain height above the Ultrahaptics. Once the user has positioned their hand there, the haptic feedback starts. However, at that moment, the button has not yet been pressed. For this to happen, the user has to move the hand down a certain distance (which simulates a physical button) and, only then, the button is pressed and a musical note sounds. When asked, 42.9% of participants stated that haptic feedback did not help at all when pushing the AR buttons and 42.9% stated that it helps to some extent. Participants liked most the fact that it was a fun game and that it is something innovative. However, most users have expressed difficulties when operating with HARptic buttons. A particular user indicated that, when his hand approached the button's position, he began to receive haptic feedback, so he thought it was already pressed. However, it was not like that, and he discovered that he had to push a little more than he found intuitive. Although other users have not mentioned it, the same behavior has been observed. As soon as they receive the haptic feedback, they assumed that the button wars already pressed.

Part of the problems that users have had when pressing the buttons is due to the following. Most of them tend to press the buttons quickly and forcefully, approaching too closely to the Ultrahaptics surface, thus complicating the hand's position tracking. This causes problems with the Leap Motion, since it is not able to correctly interpret the position of the hand (when the hand moves so fast or it leaves the working range). Those users who have started to do it slowly, tend to better manage the system. During the tests, a problem has also arisen. Once the participant received the sequence, if it takes a long time to to finish it (e.g., due to the system not properly detecting the hand), participants tend to forget the sequence and they end up pressing any random color.

### *8.5. Acceptance of AR with Mid-Air Haptics with Respect to a Standard Mobile Baseline*

Lastly, the results of Task 5 are shown here. The collaboration process seemed easy to 71.4% and intuitive to 85.9% of users. Additionally, 83.3% have indicated that AR touch mobile interaction is intuitive (during the iOS stage of the test). Even though there is a slight tie over which version seems the most intuitive to users (57.1% iOS vs. 42.9% haptics), there is an absolute agreement when asked about ease of use: 100% of users have indicated that the mobile version is easier for them to use and that it is also preferred in terms of efficiency. Finally, 85.7% of users do prefer the mobile version, while the remaining 14.3% prefer the haptic version (because it seems easier to remember physical positions than colors or they found it funnier than the iOS version).

### *8.6. Discussion and Design Notes*

From the validation with users, the following conclusions can be drawn. The evaluation of Tasks 1–3 has been done using the *high-fidelity* representation based on volume in order to offer as much detail as possible to the user. Therefore, the related result will be conditioned by both the representation and the hardware limitations. First, regarding the *interpretation* of 3D figures in the mid-air context, the concept of *inspecting* seems not to have the same meaning for all participants. Some participants understood that they had to put their hand over the Ultrahaptics and to move it up and down, forming a mental image of the 3D figure formed by 2D horizontal slices. However, other users tried to use their hand to explore the faces of the figures. With all the participants sharing the same background and the same task statement, this discrepancy attracts attention. When developing future *Harptic* systems (Haptic + AR) in which users have to explore the shape of a 3D model, in order to improve the process, it is advisable to previously indicate to the user the correct way to proceed. Second, almost all participants stated during Task 1 that haptic feedback itself was not enough to guess 3D models. However, during the course of Task 2, it has been shown that participants actually performed well during the divination process. A set of figures that seem more easily interpretable by users has been identified (*torus, house, cube, cone, and arrow*). Third, we have found that support through haptic feedback on relatively complicated AR tasks (such as establishing a certain height based on visual estimates) helps during the process. When working with interaction metaphors (as in the case of the Simon's buttons), it seems not useful to imitate an operation of the real world through haptic feedback. Therefore, when developing future metaphors, they should focus on user perception and not on replicating physical procedures known by the user. Throughout the testing process, the limitations imposed by the hardware have been present. Some users, for example, had problem because they moved their hand so fast or out of range that the integrated Leap Motion was not able to properly detect it. When taking into consideration some specific parameters, such as inspection times or resizing task accuracy, some users performed differently than others. It is difficult to determine if this is due to a single cause (e.g., previous experience on AR, inspection mode, etc.), but it is true that there may be hidden factors affecting the users acceptance.

## 9. Conclusions and Future Work

In this paper, we have presented our HARP system to enhance augmented reality perception by means of *harptic* elements, i.e., by adding haptic representation to AR elements. As a result of the whole process, we have generated a set of libraries and artifacts ready to be incorporated into future developments, which will speed up the process. At the moment, these artifacts support systems that handle Unity platform and iOS devices. However, by following the established guidelines, it is easy to adapt the HARP architecture to other platforms. Besides that, the possibility of integrating new metaphors of interaction based on haptic feedback is also a promising field. In this paper, we have explored and made use of a literal definition of *haptic representation*, focusing the process on the mid-air context. However, it is possible to integrate another type of haptic feedback such as discrete vibration. One possible application of this would be to provide haptic feedback through devices such as smartphones or smartwatches, which are increasingly more accessible. To do this, it would be enough to update the logic of the *haptic connector* component on the clients, and we are already taking steps in this direction.

Moreover, both prototypes presented in this paper are limited to only one haptic device (Ultrahaptics UHDK5). Nevertheless, the HARP architecture allows adding more instances of the *Haptic client*, each one of them associated with its own haptic device and with its own implementation. This opens the door to future scenarios in which various haptic devices can cooperate. An example of this may be, for example, that while a user is receiving feedback from Ultrahaptics, more external information is provided through haptic clues by a smartwatch.

The main objective of the user tests is to understand how haptic works to complement interactive systems based on AR. In this paper, we have focused on the users' performance when interacting with a haptic, mid-air device such as the Ultrahaptics. This implies a bias, since both prototypes used for the tests have the limitations associated with this type of system.

Facing the future, there are two pillars we are working on. First, we are currently working on integrating this architecture into an bigger, external one, which is focused on collaboration between users in AR environments. Thanks to this integration, it will be possible to haptically augment arbitrary AR content to applications and systems that were not developed for that purpose. Second, we are considering the integration of more haptic devices, in this case, wristbands and smartphones. In addition to the use cases highlighted in this paper, we believe they can be used for more advanced interaction metaphors. A good example of this would be, for example, to use them to transmit texture information to the user. If the model to be inspected has a smooth texture, the haptic feedback could be imperceptible. However, if roughness is present, its degree could be transmitted by the vibration of the device.

**Author Contributions:** Both authors designed the prototypes and experiments in the paper, analysed the data and contributed to the writing. D.V.-M. implemented the technical layer and conducted the user tests, while A.M.B. defined the initial research idea.

**Acknowledgments:** This work was supported by UPM Project RP150955017 and by the Spanish Ministry of Economy and Competitiveness under grant TEC2017-88048-C2-1-R. Additionally, the authors want to thank the trial's participants and Verónica Ruiz Bejerano for her technical help.

**Conflicts of Interest:** The authors declare no conflict of interest.